\pdfoutput=1
\documentclass[sigconf,nonacm,balance=false]{acmart}
\AtBeginDocument{}

\usepackage{babel}
\usepackage{amsmath}
\usepackage{enumitem}
\allowdisplaybreaks

\usepackage{amssymb}
\usepackage{mathtools}
\usepackage{xspace}
\usepackage{booktabs}
\usepackage{multirow}
\usepackage{fancyvrb}
\usepackage{tikz}
\usetikzlibrary{positioning,arrows.meta}
\usepackage{algorithm}
\usepackage[noend]{algpseudocode}
\definecolor{nodefill}{RGB}{232, 241, 252}
\definecolor{nodedraw}{RGB}{60, 90, 140}
\definecolor{edgecol}{RGB}{70, 70, 70}
\definecolor{lblcol}{RGB}{25, 55, 120}
\definecolor{cyclecol}{RGB}{170, 55, 55}
\definecolor{probcol}{RGB}{120, 120, 120}

\newcommand{\sys}{\ensuremath{\textsc{SPARQL}_{\mathrm{circ}}}\xspace}
\newcommand{\frag}{\ensuremath{\mathcal{S}}\xspace}

   \newcommand{\Npos}{\ensuremath{\mathbb{N}[X]}}           
          \newcommand{\BProv}[2]{\ensuremath{\mathsf{BProv}_{#1}(#2)}}
\newcommand{\cand}[2]{\ensuremath{\mathsf{Cand}_{#1}(#2)}}

\newcommand{\val}[1]{\ensuremath{\mathsf{val}_{#1}}}
\newcommand{\adom}{\ensuremath{\mathsf{adom}}}
\newcommand{\relpath}[2]{\ensuremath{\mathsf{Rel}_{#1}(#2)}}

\newcommand{\opensem}{\mathopen{[\mkern-3.2mu[}}
\newcommand{\closesem}{\mathclose{]\mkern-3.2mu]}}
\newcommand{\sem}[2]{\ensuremath{\opensem #1 \closesem_{#2}}} \newcommand{\dec}[1]{\ensuremath{\opensem #1 \closesem}}

\newcommand{\supp}{\ensuremath{\mathsf{supp}}}

\newcommand{\kw}[1]{\ensuremath{\text{\textsc{#1}}}}

\newcommand{\UNION}{\kw{union}}
\newcommand{\OPT}{\kw{optional}}
\newcommand{\MINUS}{\kw{minus}}

\newcommand{\SELECT}{\kw{select}}
\newcommand{\CONSTRUCT}{\kw{construct}}
\newcommand{\INSERT}{\kw{insert}}
\newcommand{\VALUES}{\kw{values}}
\newcommand{\WHERE}{\kw{where}}
\newcommand{\FILTER}{\kw{filter}}
\newcommand{\BIND}{\kw{bind}}
\newcommand{\AS}{\kw{as}}

\newcommand{\tc}[1]{\ensuremath{\textsc{#1}}}
\newcommand{\tp}{\mathit{tp}}
\newcommand{\opjoin}{\tc{Join}}
\newcommand{\opunion}{\tc{Union}}
\newcommand{\opdiff}{\tc{Diff}}
\newcommand{\opfilter}{\tc{Filter}}
\newcommand{\opproj}{\tc{Proj}}

\newcommand{\opopt}{\tc{Opt}}
\newcommand{\opminus}{\tc{Minus}}

\newcommand{\var}{\mathrm{var}}             \newcommand{\qtriple}[3]{\langle\!\langle #1\;#2\;#3\rangle\!\rangle}

\newcommand{\rewr}{\ensuremath{\gamma}}          \newcommand{\rewrwit}{\ensuremath{\gamma_{\mathrm{wit}}}}

\definecolor{revbcol}{rgb}{0.0,0.45,0.10}      
\definecolor{revccol}{rgb}{0.85,0.42,0.0}    
\definecolor{revdcol}{rgb}{0.80,0.0,0.60}

\definecolor{revecol}{rgb}{0.0,0.50,0.60}

\definecolor{revfcol}{rgb}{0.55,0.27,0.07}

\definecolor{revgcol}{rgb}{0.0,0.45,0.25}

\definecolor{revhcol}{rgb}{0.40,0.15,0.75}

\newcommand{\tq}[1]{\texttt{\textquotesingle#1\textquotesingle}}
\newcommand{\tseq}{\mathbin{\tq{/}}}
\newcommand{\talt}{\mathbin{\tq{|}}}
\newcommand{\tinv}{\mathord{\tq{\textasciicircum}}}

\newcommand{\plbl}[2]{{\footnotesize\bfseries\color{lblcol}$\boldsymbol{#1}$}\;{\footnotesize\bfseries\color{probcol}(#2)}}

\newcommand{\dom}{\mathrm{dom}}

       \newcommand{\reifb}[2]{\mathit{reif}(#1,#2)}   \newcommand{\rwp}[2]{\mathit{rw}(#1,#2)}
\newcommand{\grp}[1]{\texttt{\{}\,#1\,\texttt{\}}}

\newcommand{\qrows}[1]{\par\noindent
  $\begin{array}[t]{@{}l@{}}#1\end{array}$\par}
\newcommand{\qdef}[1]{\par\noindent
  $\begin{array}[t]{@{}l@{\ }l@{\ }l@{}}#1\end{array}$\par}

\AtBeginDocument{\theoremstyle{remark}}

\makeatletter\newcommand{\noproofindent}{\def\@proofindent{\noindent}}\makeatother

\begin{document}

\title[\texorpdfstring{\sys}{SPARQLcirc}: Probabilistic Query Evaluation with Provenance Circuits]{\texorpdfstring{\sys}{SPARQLcirc}:
Probabilistic Query \texorpdfstring{Evaluation}{Evaluation} with Provenance Circuits}

\author{Jingcheng Wu}
\orcid{0009-0006-1393-4760}
\affiliation{
  \institution{University of Stuttgart}
  \city{Stuttgart}
  \country{Germany}}
\email{jingcheng.wu@ki.uni-stuttgart.de}

\author{Ratan Bahadur Thapa}
\affiliation{
  \institution{University of Stuttgart}
  \city{Stuttgart}
  \country{Germany}}
\email{ratan.thapa@ki.uni-stuttgart.de}

\author{Daniel Hern\'andez}
\affiliation{
  \institution{University of Stuttgart}
  \city{Stuttgart}
  \country{Germany}}
\email{daniel.hernandez@ki.uni-stuttgart.de}

\author{Steffen Staab}
\affiliation{
  \institution{University of Stuttgart}
  \city{Stuttgart}
  \country{Germany}}
\additionalaffiliation{
  \institution{University of Southampton}
  \city{Southampton}
  \country{United Kingdom}}
\email{steffen.staab@ki.uni-stuttgart.de}

\begin{abstract}
Probabilistic query evaluation (PQE) computes the probability of every answer of a query over a tuple-independent probabilistic knowledge graph. We study exact PQE for SPARQL over graphs accessed through SPARQL~1.1 endpoints. Shared triples among the derivations of an answer make exact PQE \#P-hard in general. Knowledge compilation of the Boolean provenance of every answer yields exact probabilities, and the provenance formulas of different answers can share subformulas. We propose SPARQLcirc, which rewrites a SPARQL query into CONSTRUCT queries that name every gate by its content, so that the results of independent queries form one Boolean provenance circuit for all answers. The rewriting covers joins, unions, projection, filters, OPTIONAL, difference and transitive closures. SPARQLcirc compiles the circuit and computes the exact probability of every answer by weighted model counting. For the evaluation, we draw random probabilities for the triples of WatDiv and gMark and the statements of Wikidata and evaluate the CONSTRUCT queries on GraphDB, Fuseki and Oxigraph. The ratio of the end-to-end time of PQE with SPARQLcirc to that of PQE by compiling the provenance expressions of the SPARQL provenance rewriting NPCS decreases with the number of derivations.
\end{abstract}



\maketitle

\section{Introduction}\label{sec:intro}
\begin{figure}[htbp]
\centering
\begin{tikzpicture}[
    drug/.style={draw=nodedraw, fill=nodefill, rounded corners,
                 font=\small\bfseries,
                 minimum width=1.4cm, minimum height=0.55cm},
    arr/.style={->, >=stealth, semithick, edgecol},
    cyc/.style={->, >=stealth, semithick, dashed, cyclecol},
    lbl/.style={fill=white, inner sep=1pt},
    yscale=0.85
]
  \node[drug] (war) at (2.7, -1) {Warfarin};
  \node[drug] (clo) at (2, 0) {Clopidogrel};
  \node[drug] (ome) at (5.5, -2.5) {Omeprazole};
  \node[drug] (asp) at (0, -1.6) {Aspirin};
  \node[drug] (lis) at (5.5, 0) {Lisinopril};
  \node[drug] (met) at (6, -1.1) {Metformin};
  \node[drug] (ibu) at (2.7, -2.3) {Ibuprofen};
  \draw[cyc] (asp) to[bend left=15] node[lbl] {\plbl{p_1}{0.92}} (war.180);
  \draw[arr] (war.270) to[bend right=15] node[lbl] {\plbl{p_2}{0.87}} (met.180);
  \draw[arr] (met) to[bend left=15] node[lbl] {\plbl{p_3}{0.85}} (ome.90);
  \draw[arr] (asp) to[bend right=15] node[lbl] {\plbl{p_4}{0.78}} (ibu.180);
  \draw[arr] (ibu) to[bend right=15] node[lbl] {\plbl{p_5}{0.71}} (met.180);
  \draw[cyc] (war.360) to[bend right=10] node[lbl] {\plbl{p_6}{0.65}} (lis.270);
  \draw[cyc] (lis) -- node[lbl] {\plbl{p_7}{0.60}} (clo);
  \draw[cyc] (clo.180) to[bend right=25] node[lbl] {\plbl{p_8}{0.55}} (asp);
  \draw[arr] (0.1, -3.15) -- (0.8, -3.15);
  \node[anchor=west, font=\footnotesize] at (0.85, -3.15) {triple off the cycle};
  \draw[cyc] (3.9, -3.15) -- (4.6, -3.15);
  \node[anchor=west, font=\footnotesize] at (4.65, -3.15) {triple on the cycle};
\end{tikzpicture}
\caption{Probabilistic drug-interaction graph. Every edge is an \texttt{:interactsWith} triple with its identifier and probability.}
\Description{A directed graph of seven drugs connected by eight interactsWith edges, labeled with the identifiers p1 to p8 and their probabilities. The dashed edges p1, p6, p7 and p8 form the only directed cycle, through Aspirin, Warfarin, Lisinopril and Clopidogrel. A legend below the graph explains the solid and the dashed edges.}
\label{fig:running}
\end{figure}
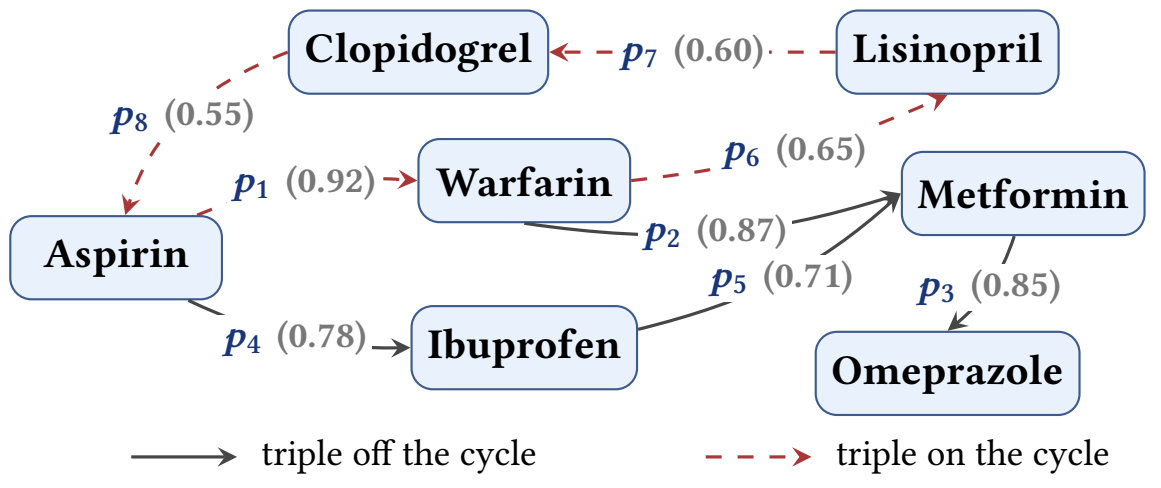

Information extraction and link prediction produce triples with confidence scores~\cite{etzioni2004knowitall,carlson2010coupled,dong2014knowledge,chen2019ukg}, and statistical signal detection derives drug interactions with association scores from adverse-event reports~\cite{tatonetti2012twosides}. A probabilistic knowledge graph treats every score as the probability that the triple holds~\cite{ceylan2016openworld}. In RDF, the probability annotates the triple through standard reification, named graphs, RDF-star or Wikidata statement nodes~\cite{w3c-rdfschema11,carroll2005namedgraphs,hartig2021rdfstar,rdf12concepts,malyshev2018wikidata}. In Figure~\ref{fig:running}, every \texttt{:interactsWith} triple carries an identifier $p_i$ and a probability. Knowledge graphs with annotated triples, such as Wikidata, are queried through SPARQL endpoints~\cite{vrandecic2014wikidata,malyshev2018wikidata}. A SPARQL query can return the probabilities of the matched triples, but the SPARQL semantics assigns no probability to an answer~\cite{w3c-sparql11}. A public SPARQL endpoint accepts only SPARQL requests, and the endpoint administrator alone can extend the query engine. For a probabilistic knowledge graph accessed only through a public endpoint on the Web, the probability of every answer must therefore be computed from the results of SPARQL queries.

In the tuple-independent model of probabilistic databases, every subgraph of a probabilistic knowledge graph is a \emph{possible world}, and every triple occurs independently with its probability~\cite{suciu2011probdb,ceylan2016openworld}. The \emph{marginal probability} of an answer is the total probability of the worlds in which the answer occurs. \emph{Probabilistic query evaluation} (PQE) computes the marginal probability of every answer. Top-k query evaluation ranks the answers by marginal probability~\cite{re2007topk}. The \emph{Boolean provenance} of an answer is the Boolean function over the triples that is true exactly in the worlds in which the answer occurs, so the marginal probability is the probability that the Boolean provenance is true~\cite{suciu2011probdb}. The identifiers $p_i$ of Figure~\ref{fig:running} also denote the Boolean variables of their triples. The query
{\begin{verbatim}
  SELECT ?z WHERE { :Aspirin :interactsWith ?x .
    ?x :interactsWith ?y . ?y :interactsWith ?z }
\end{verbatim}}
\noindent retrieves the drugs that Aspirin reaches in three \texttt{:interactsWith} steps. The Boolean provenance is $p_1\land p_6\land p_7$ for Clopidogrel and $(p_1\land p_2\land p_3)\lor(p_4\land p_5\land p_3)$ for Omeprazole. A \emph{derivation} of an answer is the conjunction of the triples of one match. Omeprazole has two derivations and Clopidogrel one.

Derivations that share triples are correlated events. The two derivations of Omeprazole have probabilities $0.680$ and $0.471$ and both contain $p_3$. Treating the derivations as independent events gives $1-(1-0.680)(1-0.471)=0.831$, whereas factoring out $p_3$ gives the correct marginal probability $0.85\cdot(1-(1-0.800)(1-0.554))=0.774$. Even for some fixed basic graph patterns, exact PQE is \#P-hard in the size of the graph~\cite{dalvi2012dichotomy,valiant1979}. Knowledge compilation translates the Boolean provenance into a compiled form, from which \emph{weighted model counting} computes the marginal probability in time linear in the size of the form~\cite{darwiche2002kcmap,suciu2011probdb}. A \emph{provenance circuit} is a directed acyclic graph that represents Boolean provenance with triples as leaves and with products ($\land$), sums ($\lor$) and differences ($a\land\lnot b$) as gates~\cite{deutch2014circuits,senellart2018provsql}. Derivations of different answers can also share triples and subformulas. A single circuit represents a shared subformula once and gives every answer a \emph{root} gate~\cite{deutch2014circuits,senellart2018provsql}.

SPARQL property paths with the operator \texttt{+} or \texttt{*} compute transitive closures. On a graph with a cycle, a closure can connect two nodes by infinitely many walks. In Figure~\ref{fig:running}, the walks of \texttt{:Aspirin :interactsWith+ ?y} from Aspirin to Warfarin can repeat the cycle. Provenance that counts the walks becomes a power series with infinitely many terms~\cite{green2007provenance}, and the series for Warfarin is $p_1+p_1^2p_6p_7p_8+\cdots$. In contrast, by the absorption law $x\lor(x\land y)=x$, the Boolean provenance of every answer equals a disjunction over finitely many walks~\cite{deutch2014circuits}. The law removes the conjunction of every walk that contains all triples of a shorter walk between the same nodes, and for Warfarin the disjunction $p_1\lor(p_1\land p_6\land p_7\land p_8)\lor\cdots$ equals $p_1$. For transitive closure, positive provenance formulas require super-polynomial size in the worst case, whereas polynomial-size circuits exist for every graph~\cite{deutch2014circuits,fan2025circuits}.

SPARQLprov and NPCS rewrite a query into SPARQL queries, so that an endpoint without extensions returns the provenance of every answer~\cite{hernandez2021sparqlprov,asma2026npcs}. SPARQLprov returns tables that encode expression structure and node identifiers. NPCS constructs and aggregates expression strings inside the endpoint. Both rewritings return one provenance expression per answer, and a subexpression that several answers share is therefore repeated in each expression. The probability of every answer is then computed by compiling its expression outside the endpoint. A SPARQL property path returns only the pairs of nodes that its walks connect, and SPARQLprov and NPCS cover only patterns without transitive closures. In relational databases, ProvSQL covers recursive reachability queries, builds one circuit for all answers and names every inner gate by hashing its operator and operands~\cite{senellart2018provsql,sen2025provsql,provsql2026recursion}. ProvSQL runs inside PostgreSQL as an engine extension, but engine extensions on a public endpoint can be installed only by its administrator.

SPARQL \CONSTRUCT{} queries can return the gates of a circuit as triples, but a blank node in the template denotes a fresh node for every solution. Two gates that different solutions or queries emit therefore coincide only if both receive the same IRI. The \emph{name} of a gate is a canonical syntactic description of the gate. A SPARQL~1.1 query can hash the name of a gate into its IRI with built-in hash functions, and equal names yield equal IRIs in every query and on every engine. The results of independent \CONSTRUCT{} queries then form one circuit, because the triples that two queries emit for one gate share the IRI of the gate. Repeated edges occur once in the merged RDF graph, and by Boolean idempotence ($a\lor a=a$ and $a\land a=a$) the merge preserves the Boolean function of every gate.

We propose \sys, a rewriting method for exact PQE. \sys translates a SPARQL query into a \emph{plan} of SPARQL~1.1 \CONSTRUCT{} queries whose results form one Boolean provenance circuit for all answers. For a transitive closure from a start node that a constant or a preceding pattern fixes, each round of SPARQL~1.1 requests emits the gates of one walk length. \sys compiles the root of every answer with an existing knowledge compiler and computes the probability of the answer by weighted model counting. With random probabilities for the triples of WatDiv~\cite{aluc2014watdiv}, gMark~\cite{gmark} and the 1.78 billion statements of Wikidata~\cite{vrandecic2014wikidata}, we compare \sys on GraphDB~\cite{graphdb}, Fuseki~\cite{fuseki} and Oxigraph~\cite{oxigraph} with plain query evaluation and NPCS. We make three contributions:
\begin{itemize}[leftmargin=1.2em,itemsep=1pt,parsep=0pt,topsep=2pt]
\item Gate names that SPARQL~1.1 queries compute from the content of the gates, so that independent \CONSTRUCT{} queries yield one circuit on any engine (Section~\ref{sec:names}).
\item A rewriting of queries with \OPT{}, difference and transitive closures into plans of \CONSTRUCT{} queries, a \emph{factored plan} in which derivations share the gates of common parts, and a proof that every answer has a root and that the root denotes its Boolean provenance (Sections~\ref{sec:rewriting}--\ref{sec:correctness}).
\item An evaluation on WatDiv, gMark and Wikidata in which the ratio of the end-to-end time of PQE with one circuit for all answers to that with per-answer provenance expressions decreases with the number of derivations (Section~\ref{sec:eval}).
\end{itemize}
Section~\ref{sec:related} relates \sys to probabilistic relational and RDF databases, SPARQL provenance and provenance under recursion.
\section{Preliminaries}\label{sec:prelim}

\noindent\textbf{RDF and SPARQL.}
Following Pérez et al.~\cite{perez2009semantics}, we define a fragment \frag of SPARQL as follows. The set $\mathbf{I}$ contains the IRIs, $\mathbf{B}$ the blank nodes, $\mathbf{L}$ the literals and $\mathbf{V}$ the variables, and $\mathbf{T}=\mathbf{I}\cup\mathbf{B}\cup\mathbf{L}$. An RDF graph is a finite subset of $(\mathbf{I}\cup\mathbf{B})\times\mathbf{I}\times\mathbf{T}$, a triple pattern $\tp$ belongs to $(\mathbf{I}\cup\mathbf{L}\cup\mathbf{V})\times(\mathbf{I}\cup\mathbf{V})\times(\mathbf{I}\cup\mathbf{L}\cup\mathbf{V})$, and a solution mapping is a partial function $\mu:\mathbf{V}\rightharpoonup\mathbf{T}$ with domain $\dom(\mu)$. A blank node in a query acts as a variable~\cite{w3c-sparql11}. Mappings $\mu_1$ and $\mu_2$ are compatible ($\mu_1\sim\mu_2$) when both agree on $\dom(\mu_1)\cap\dom(\mu_2)$. A path-free pattern is generated by
\[
 \begin{aligned}
  P ::={}& {\tp} \mid \opjoin(P_1,\ldots,P_n) \mid \opunion(P_1,\ldots,P_n)\\
        {}&\mid \opproj_{Y}(P) \mid \opfilter_\varphi(P) \mid \opdiff(P_1,P_2).
 \end{aligned}
\]
A basic graph pattern (BGP) $\{\tp_1,\ldots,\tp_m\}$ is represented by $\opjoin(\tp_1,\ldots,\tp_m)$~\cite{perez2009semantics,w3c-sparql11}, so the empty BGP $\opjoin()$ returns only the empty mapping. The condition $\varphi$ combines Boolean connectives, comparisons and arithmetic operators with the SPARQL~1.1 built-ins of Appendix~\ref{app:builder}. In $\opfilter_\varphi(P)$, the condition $\varphi$ mentions only variables that every mapping of $P$ binds. A query is a pattern $Q=\opproj_Y(P)$ with $Y\subseteq\mathbf{V}$. Evaluation of $P$ on an RDF graph $G$ returns a multiset $\sem{P}{G}$~\cite{angles2016multiset,w3c-sparql11}, and $\supp(\sem{P}{G})$ denotes the mappings with positive multiplicity. The difference $\opdiff(P_1,P_2)$~\cite{kontchakov2016expressibility} retains a mapping of $P_1$ exactly when no mapping of $P_2$ is compatible with it, and $\opopt(P_1,P_2)$ without a filter abbreviates $\opunion(\opjoin(P_1,P_2),\opdiff(P_1,P_2))$~\cite{w3c-sparql11}.

A closure-free property-path expression is generated by $e ::= p \mid \tinv e \mid e_1\tseq e_2 \mid e_1\talt e_2$, with $p\in\mathbf I$ (inverse, sequence and alternative). $\relpath{e}{G}$ is the binary relation that SPARQL assigns to $e$ on $G$~\cite{w3c-sparql11}, and $\relpath{e^+}{G}$ is its transitive closure. For $\diamond\in\{+,*\}$, a closure atom is a pattern $C=(\alpha,e^\diamond,?y)$ with closure-free $e$, source $\alpha\in\mathbf{I}\cup\mathbf{L}\cup\mathbf{V}$ and target variable $?y$. For a source $s$ and a target $v$, the mapping $\mu^C_{s,v}$ is $\{?y\mapsto v\}$ when $\alpha$ is a constant and $\{\alpha\mapsto s,?y\mapsto v\}$ when $\alpha$ is a variable. For $\diamond=+$, the atom returns $\mu^C_{s,v}$ once for every pair $(s,v)\in\relpath{e^+}{G}$ with $s=\alpha$ when $\alpha$ is a constant and $s=v$ when $\alpha$ is $?y$. For $\diamond=*$, the atom returns $\mu^C_{s,v}$ once for every such pair of $\relpath{e^+}{G}\cup \Delta$, where the set $\Delta$ of zero-length pairs contains $(\alpha,\alpha)$ when $\alpha$ is a constant and $(s,s)$ for every subject or object $s$ of a triple of $G$ when $\alpha$ is a variable~\cite{w3c-sparql11}. A closure atom $C$ with a variable source is \emph{source-bound} when $C$ is an operand of a join or the right operand of a difference and, on every graph, every mapping of the join of the operands before $C$ binds the source. The fragment \frag extends this grammar with closure atoms whose source is a constant or that are source-bound.

A \CONSTRUCT{} query pairs a pattern $P$ with a template $T$, a finite set of triple patterns. On an RDF graph $G$, we call the mappings in $\supp(\sem{P}{G})$ the \emph{rows} of the query, and the query returns the RDF graph that contains the instance $\mu(T)$ of every row $\mu$. The instance $\mu(T)$ consists of the RDF triples among the triples that result from $T$ when every variable of $\dom(\mu)$ is replaced by its value~\cite{w3c-sparql11}.

\noindent\textbf{Reification.}
Four encodings give every triple $t=(s,p,o)$ of an RDF graph $G$ an identifier $i$: an RDF-star annotation $\qtriple{s}{p}{o}\ \mathrm{occurrenceOf}\ i$~\cite{hartig2021rdfstar}, a standard reification $i$~\cite{w3c-rdfschema11}, a named graph $i$ that holds $t$~\cite{carroll2005namedgraphs}, or a Wikidata statement node $i$ of best rank~\cite{malyshev2018wikidata}. SPARQL-star queries read the RDF-star annotation through quoted triple patterns~\cite{hartig2021rdfstar}, and SPARQL~1.1 queries read the other three encodings. The reified graph $G_{\mathrm{reif}}$ replaces every blank node of $G$ by a fresh Skolem IRI~\cite{w3c-rdf11}, contains the resulting triples and represents each of them in one of the encodings. Following SPARQLprov and NPCS~\cite{hernandez2021sparqlprov,asma2026npcs}, the reification $\reifb{\tp}{?t}$ of a triple pattern $\tp$ matches the encoding of every triple that matches $\tp$ and binds $?t$ to the identifier of the triple.

\noindent\textbf{Possible worlds and PQE.}
A probabilistic knowledge graph is an RDF graph $G$ in which every triple $t$ carries a Boolean variable $x_t$ and a probability $\Pr(x_t)$~\cite{suciu2011probdb,ceylan2016openworld}. Each triple $t$ has one distinct identifier $i_t$ in $G_{\mathrm{reif}}$. Under tuple independence, a possible world $W\subseteq G$ has probability $\Pr(W)=\prod_{t\in W}\Pr(x_t)\prod_{t\in G\setminus W}(1-\Pr(x_t))$. A mapping $\mu$ is an answer of a query $Q$ when $\mu\in\supp(\sem{Q}{W})$ for some world $W\subseteq G$. The marginal probability of the answer is $\Pr(\mu)=\sum_{W\subseteq G,\ \mu\in\supp(\sem{Q}{W})}\Pr(W)$.

\noindent\textbf{Boolean provenance.}
Geerts et al.\ define the provenance of SPARQL over spm-semirings, which extend commutative semirings with a difference $\ominus$~\cite{geerts2016algebraic}. The homomorphism that maps $\oplus$ to $\lor$, $\otimes$ to $\land$ and $a\ominus b$ to $a\land\neg b$ yields the Boolean provenance $\BProv{P}{\mu}$ of a mapping $\mu$ for a pattern $P$. Over $X=\{x_t\mid t\in G\}$, the Boolean provenance has the following cases. For a triple pattern $\tp$, $\BProv{\tp}{\mu}=x_{\mu(\tp)}$ when $\dom(\mu)$ contains exactly the variables of $\tp$ and $\mu(\tp)\in G$, and $\bot$ otherwise. The other cases are
\begin{align*}
 \BProv{\opjoin(P_1,\ldots,P_n)}{\mu}&=\textstyle\bigvee_{\mu=\mu_1\cup\cdots\cup\mu_n}\bigwedge_{i}\BProv{P_i}{\mu_i},\\
 \BProv{\opunion(P_1,\ldots,P_n)}{\mu}&=\textstyle\bigvee_{i}\BProv{P_i}{\mu},\\
 \BProv{\opproj_{Y}(P)}{\mu}&=\textstyle\bigvee_{\mu'|_{Y}=\mu}\BProv{P}{\mu'},\\
 \BProv{\opdiff(P_1,P_2)}{\mu}&=\BProv{P_1}{\mu}\land\neg\textstyle\bigvee_{\nu\sim\mu}\BProv{P_2}{\nu},
\end{align*}
where the join ranges over decompositions into pairwise compatible mappings, and $\BProv{\opfilter_\varphi(P)}{\mu}$ is $\BProv{P}{\mu}$ when $\mu$ satisfies $\varphi$ and $\bot$ otherwise. Let $\val{W}$ denote the valuation that assigns $\top$ to $x_t$ exactly when $t\in W$. For a closure atom, $\BProv{C}{\mu^C_{s,v}}$ is the Boolean function $f$ over $X$ with $\val{W}(f)=\top$ exactly when $C$ returns $\mu^C_{s,v}$ on $W$, and $\BProv{C}{\mu}=\bot$ for every other $\mu$. For every $W\subseteq G$, the commutation of the provenance with spm-semiring homomorphisms~\cite{geerts2016algebraic} and the definition of $\BProv{C}{\cdot}$ yield
\begin{equation}
 \val{W}(\BProv{P}{\mu})=\top
 \,\,\Longleftrightarrow\,\,
 \mu\in\supp(\sem{P}{W}).
 \label{eq:world}
\end{equation}

\noindent\textbf{Provenance circuits.}
Following Deutch et al.~\cite{deutch2014circuits} and ProvSQL~\cite{senellart2018provsql}, a provenance circuit over $X$ is a finite set of gates whose child relation is acyclic. A gate $g$ is a leaf $x\in X$, a sum gate ($\oplus$) or a product gate ($\otimes$) with a finite set $\mathit{ch}(g)$ of children, or a difference gate ($\ominus$) with a finite set $\mathit{pos}(g)$ of minuend children and one subtrahend child $\mathit{neg}(g)$. ProvSQL restricts difference gates to one minuend. A partial function $\mathrm{root}$ maps solution mappings to gates, and $\mathrm{root}(\mu)$ is the root of $\mu$. By recursion over the child relation, the Boolean semantics $\dec{g}$ of a gate is $x$ for a leaf $x$, $\bigvee_{c\in\mathit{ch}(g)}\dec{c}$ for an $\oplus$-gate, $\bigwedge_{c\in\mathit{ch}(g)}\dec{c}$ for an $\otimes$-gate, and
\begin{equation}
 \dec{g}=\Bigl(\bigvee_{c\in\mathit{pos}(g)}\dec{c}\Bigr)\land\neg\dec{\mathit{neg}(g)}
 \label{eq:circuit-semantics}
\end{equation}
for an $\ominus$-gate, so $g_\bot=\oplus(\varnothing)$ and $g_\top=\otimes(\varnothing)$ denote $\bot$ and $\top$.

\noindent\textbf{Knowledge compilation and weighted model counting.}
The weighted model count $\mathrm{WMC}(f)$ of a Boolean function $f$ sums over the satisfying assignments of $f$ the product of the weights of their literals~\cite{sang2005wmc}. With the weights $\Pr(x_t)$ on $x_t$ and $1-\Pr(x_t)$ on $\neg x_t$, Equation~\eqref{eq:world} yields $\Pr(\mu)=\mathrm{WMC}(\BProv{Q}{\mu})$. Knowledge compilation translates a Boolean function into a deterministic decomposable negation normal form (d-DNNF) or an ordered binary decision diagram (OBDD), and every OBDD is a d-DNNF~\cite{darwiche2002kcmap}. A d-DNNF is a circuit in which negation occurs only on variables, the children of every conjunction have pairwise disjoint sets of variables and the children of every disjunction are mutually exclusive~\cite{darwiche2002kcmap}. With the weights above, the weighted model counts of the children therefore multiply at a conjunction and add at a disjunction, and one bottom-up pass computes the weighted model count of a d-DNNF in linear time. The d-DNNF
\[
  p_3\land\bigl((p_1\land(p_2\lor(\neg p_2\land p_4\land p_5)))\lor(\neg p_1\land p_4\land p_5)\bigr)
\]
represents the Boolean provenance of Omeprazole, and the pass yields the marginal probability $0.774$.
\section{Problem Definition}\label{sec:problem}
{\setlength{\parindent}{0pt}\begin{definition}[Problem]\label{def:problem}
Given a query $Q$ of \frag and the probability $\Pr(x_t)$ for the identifier $i_t$ of every $t\in G$, the problem is to compute $\Pr(\mu)$ for every answer $\mu$ of $Q$ from the results of SPARQL~1.1 requests, or SPARQL-star requests for the RDF-star annotation, to an endpoint that stores $G_{\mathrm{reif}}$.
\end{definition}}

By Equation~\eqref{eq:world}, $\Pr(\mu)$ equals $\mathrm{WMC}(\BProv{Q}{\mu})$, so it suffices that the results of such requests form a provenance circuit in which the root of every answer $\mu$ denotes $\BProv{Q}{\mu}$.
\section{\texorpdfstring{Provenance Circuits from SPARQL~1.1 Queries}{Provenance Circuits from SPARQL 1.1 Queries}}\label{sec:method}
In this section, we rewrite a query into a \emph{plan} of SPARQL~1.1 \CONSTRUCT{} queries whose results form one Boolean provenance circuit for all answers. Let $\tp_1$, $\tp_2$ and $\tp_3$ be the triple patterns of the running query in their order of occurrence. The rewriting $\rewr$ maps the algebra expression $\opproj_{\{?z\}}(\opjoin(\tp_1,\tp_2,\tp_3))$ to a plan with one query for the join and one for the projection. The two queries share the \WHERE{} clause and appear as one query below.

\addvspace{3pt}\noindent\begin{minipage}[c]{0.9\columnwidth}
\begin{Verbatim}[fontsize=\footnotesize,frame=single,framesep=3pt,commandchars=\|\[\]]
CONSTRUCT { ?g a c:Times ; c:in ?t1, ?t2, ?t3 .
            ?r a c:Plus ; c:in ?g ;
               c:answerRoot "vars:z" ; c:bind:z ?z }
WHERE { << :Aspirin :interactsWith ?x >> :occurrenceOf ?t1 .
        << ?x :interactsWith ?y >> :occurrenceOf ?t2 .
        << ?y :interactsWith ?z >> :occurrenceOf ?t3 .
        BIND(KEY_TIMES(?t1, ?t2, ?t3) AS ?g)
        BIND(KEY_PLUS(A, "Proj{z}(...)", ?z) AS ?r) }
\end{Verbatim}
\end{minipage}\hfill\refstepcounter{equation}\label{eq:plan}(\theequation)\par
\addvspace{3pt}

The shared \WHERE{} clause matches the reifications of the three triple patterns and binds the identifiers \texttt{?t1}, \texttt{?t2} and \texttt{?t3} of the matched triples. The two \BIND{} expressions name the product gate \texttt{?g} and the root \texttt{?r} by \texttt{KEY\_TIMES} and \texttt{KEY\_PLUS} of Definition~\ref{def:keys}. For every row, the template emits one triple per gate for its kind (\texttt{c:Times} or \texttt{c:Plus}) and one \texttt{c:in} triple per edge to a child. The last template line marks \texttt{?r} as a root and records the value of \texttt{?z}.

\begin{figure}[htbp]
\centering
\begin{tikzpicture}[
  >=stealth, every node/.style={font=\normalsize\bfseries},
  ans/.style={draw, rounded corners, fill=gray!12, inner sep=2pt,
              minimum height=5mm, font=\normalsize\bfseries},
  gate/.style={draw, circle, minimum size=7mm, inner sep=0pt,
               font=\normalsize},
  leaf/.style={draw, fill=blue!8, minimum size=6mm, inner sep=1pt},
  shleaf/.style={draw, fill=orange!18, minimum size=6mm, inner sep=1pt},
  scale=0.8, transform shape
]
  \node[ans] (aOme) at (2.2,3.0) {Omeprazole};
  \node[ans] (aClo) at (5.8,3.0) {Clopidogrel};
  \node[gate] (gO) at (2.2,1.9) {$\oplus$};
  \node[gate] (gCp) at (5.8,1.9) {$\oplus$};
  \node[gate] (gC) at (5.8,0.9) {$\otimes$};
  \node[gate] (gB) at (1.1,0.9) {$\otimes$};
  \node[gate] (gA) at (3.3,0.9) {$\otimes$};
  \node[leaf] (p4) at (0.0,0.0) {$p_4$};
  \node[leaf] (p5) at (1.1,0.0) {$p_5$};
  \node[shleaf] (p3) at (2.2,0.0) {$p_3$};
  \node[leaf] (p2) at (3.3,0.0) {$p_2$};
  \node[shleaf] (p1) at (4.4,0.0) {$p_1$};
  \node[leaf] (p6) at (5.8,0.0) {$p_6$};
  \node[leaf] (p7) at (7.2,0.0) {$p_7$};
  \draw[->] (aOme) -- (gO);
  \draw[->] (aClo) -- (gCp);
  \draw[->] (gCp) -- (gC);
  \draw[->] (gO) -- (gB);
  \draw[->] (gO) -- (gA);
  \draw[->] (gB) -- (p4);
  \draw[->] (gB) -- (p5);
  \draw[->] (gB) -- (p3);
  \draw[->] (gA) -- (p3);
  \draw[->] (gA) -- (p2);
  \draw[->] (gA) -- (p1);
  \draw[->] (gC) -- (p1);
  \draw[->] (gC) -- (p6);
  \draw[->] (gC) -- (p7);
\end{tikzpicture}
\caption{The circuit that the plan of the running query emits. The shared leaves $p_1$ and $p_3$ (orange) occur once.}
\Description{The circuit of the running query: the answers Omeprazole and Clopidogrel, each above its root plus-gate, and three product gates below the roots that share the leaves p1 and p3.}
\label{fig:overview}
\label{fig:circuit}
\end{figure}
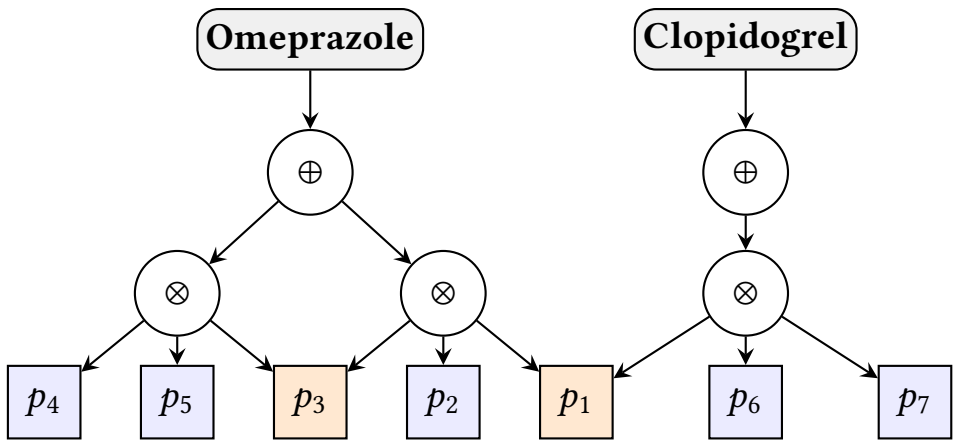

Over $G_{\mathrm{reif}}$, the plan emits the product gates $\otimes(p_1,p_2,p_3)$ and $\otimes(p_4,p_5,p_3)$ below the root of Omeprazole and the product gate $\otimes(p_1,p_6,p_7)$ below the root of Clopidogrel. The merged RDF graph of the emitted triples is the circuit of Figure~\ref{fig:overview}, and two product gates share each of the leaves $p_1$ and $p_3$. The plan depends on the query alone, and the circuit on the query and the data. \sys reads the circuit, compiles every root and computes its weighted model count: $0.774$ for Omeprazole and $0.359$ for Clopidogrel.

Section~\ref{sec:names} names the gates. Sections~\ref{sec:rewriting} to~\ref{sec:paths} give the plans for the operators, for basic graph patterns and for transitive closures, and Section~\ref{sec:correctness} defines sound and complete plans, gives the Boolean functions of the named gates, states that the plans are sound and complete and compiles the circuit to compute the probabilities. Appendix~\ref{app:proofs} proves the lemmas and theorems of this section.

\subsection{\texorpdfstring{Gate names and merging}{Gate names and merging}}\label{sec:names}
\sys merges the results of all queries of the plan into one RDF graph. All triples with one gate IRI describe one gate. An RDF graph contains every triple at most once, so an edge emitted by several rows or queries occurs once. One copy of an edge preserves the value of every gate exactly when sum and product are idempotent. Boolean functions are idempotent, whereas the provenance polynomials of $\Npos$ satisfy $a+a=2a$~\cite{green2007provenance}.

The IRI of every gate other than a leaf is the value of a \BIND{} expression, which reads one row. The row of a join binds the gates of all operands, so a product gate is named after the list of its children, in which a gate repeats when two operands match the same triple. The edges into a sum or difference gate come from the rows that carry one mapping. Each of these rows contributes one derivation, and the rows can belong to several queries. All rows of one mapping bind the same values, so a sum or difference gate is named after its subpattern and the bindings of its mapping. Definition~\ref{def:ids} adapts the gate names of ProvSQL~\cite{sen2025provsql} and the node identifiers of SPARQLprov~\cite{hernandez2021sparqlprov} to SPARQL~1.1 queries, and hash-consing names the nodes of terms in the same way~\cite{filliatre2006hashconsing}.

\begin{definition}[Gate names]\label{def:ids}\label{def:keys}
Let $\tau(v)$ encode an RDF term $v$ by its term kind, lexical form, datatype and lower-cased language tag, and let $\tau(\mathsf{unbound})$ be a distinct token for an unbound variable. Let $\theta(P)$ be a serialization of a subpattern $P$ such that two subpatterns have equal serializations exactly when they are equal up to the order of the operands of joins and unions and up to repeated operands of a union. A leaf is named by the identifier of its triple. The key expressions for a letter $L$, a tag $\theta$, variables $?c_1,\ldots,?c_n$ and variables $?x_1<\cdots<?x_k$ in the lexicographic order of their names are
\[
\begin{array}{@{}r@{\;}l@{}}
\texttt{KEY\_TIMES}(?c_1,\ldots,?c_n)&=\texttt{IRI}(h(\texttt{T},\operatorname{sort}(?c_1,\ldots,?c_n))),\\
\texttt{KEY\_PLUS}(L,\theta,?x_1,\ldots,?x_k)&=\texttt{IRI}(h(L,\theta,\hat\tau(?x_1),\ldots,\hat\tau(?x_k))),\\
\texttt{KEY\_MINUS}(\theta,?x_1,\ldots,?x_k)&=\texttt{IRI}(h(\texttt{M},\theta,\hat\tau(?x_1),\ldots,\hat\tau(?x_k))),
\end{array}
\]
where $h$ hashes its arguments, $\operatorname{sort}$ orders them by the lexicographic order of their hashes, $\hat\tau(?x)$ computes $\tau$ of the value of $?x$, and $L\in\{\texttt{U},\texttt{P},\texttt{A},\texttt{S}\}$ marks the sum gate of a union, a projection, the outer projection and the subtrahend of a difference, respectively. The \emph{name} of a gate on a row is the tuple of the arguments of $h$, and its \emph{IRI} is the value of the key expression.
\end{definition}

Since $\otimes$ is commutative but $h$ depends on the order of its arguments, \texttt{KEY\_TIMES} sorts the children before hashing, so the name of a product gate is invariant under permutations of the operands of a join. In the plan~\eqref{eq:plan}, the row that binds \texttt{?t1}, \texttt{?t2} and \texttt{?t3} to $p_1$, $p_2$ and $p_3$ names its product gate $(\texttt{T},\operatorname{sort}(p_1,p_2,p_3))$, and a query that lists the triple patterns in the order $\tp_3$, $\tp_2$, $\tp_1$ names the same gate. The root of Omeprazole is named $(\texttt{A},\theta(\opproj_{\{?z\}}(\opjoin(\tp_1,\tp_2,\tp_3))),\tau(\mathrm{Omeprazole}))$. The query $\opproj_{\{?z\}}(\tp_3)$ names the root of Omeprazole with another tag, and that root denotes $p_3$. A name without the tag merges the two roots and gives Omeprazole the function $p_3$ in the running query instead of $(p_1\land p_2\land p_3)\lor(p_4\land p_5\land p_3)$. Conversely, gates with the same Boolean function can have distinct names, because a \BIND{} expression computes a name from the values of one row, whereas the function of a gate depends on all gates below it. For example, $\opproj_{\{?z\}}(\tp_3)$ and $\opproj_{\{?z\}}(\opjoin(\tp_3,\tp_3))$ have the same Boolean provenance, but their sum gates have distinct tags.

\subsection{Path-free patterns}\label{sec:rewriting}
The rewriting $\rewr$ translates a \SELECT{} query $Q$ into a plan $\rewr(Q)$ of \CONSTRUCT{} queries, as in~\eqref{eq:plan}. For every subpattern $P$, Definition~\ref{def:gamma} defines a rewritten pattern $\rwp{P}{g_P}$ and the queries of $P$. The rows of $\rwp{P}{g_P}$ bind the mappings of $P$ and their gates. The queries of $P$ output these gates and their edges to the gates of the operands.

\begin{definition}[Rewriting]\label{def:gamma}
Let $P$ be a subpattern of a query that is not a closure atom, and let $g_P$ be a variable, called the gate variable of $P$. The rewritten pattern $\rwp{P}{g_P}$, the \CONSTRUCT{} queries of $P$ and the plan $\rewr(P)$ are defined by recursion on $P$.
\begin{enumerate}[label=(\arabic*),leftmargin=2em,itemsep=1pt,parsep=0pt,topsep=0pt,before=\raggedright]
\item If $P$ is a triple pattern $\tp$, then $\rwp{P}{g_P}=\reifb{\tp}{g_P}$ and $\rewr(P)=\varnothing$.
\item If $P=\opjoin(P_1,\ldots,P_n)$, then, for fresh variables $g_{P_1},\ldots,g_{P_n}$,
\qrows{
\rwp{P}{g_P}=\grp{\rwp{P_1}{g_{P_1}}}\cdots\grp{\rwp{P_n}{g_{P_n}}}\\
\qquad\BIND(\texttt{KEY\_TIMES}(g_{P_1},\ldots,g_{P_n})\ \AS\ g_P),
}
\qdef{
q_P= & \CONSTRUCT & \grp{g_P\ \texttt{a c:Times ; c:in}\ g_{P_1},\ldots,g_{P_n}}\\
 & \WHERE & \grp{\rwp{P}{g_P}},
}
\qrows{
\rewr(P)=\{q_P\}\cup\rewr(P_1)\cup\cdots\cup\rewr(P_n).
}
\item If $P=\opunion(P_1,\ldots,P_n)$, then, for fresh variables $g_{P_1},\ldots,g_{P_n}$, the tuple $\bar v_P$ of the variables in scope at $P$ in lexicographic order and $i\in\{1,\ldots,n\}$,
\qrows{
\rwp{P}{g_P}=\grp{\rwp{P_1}{g_{P_1}}}\ \UNION\ \cdots\ \UNION\ \grp{\rwp{P_n}{g_{P_n}}}\\
\qquad\BIND(\texttt{KEY\_PLUS}(\texttt{U},\theta(P),\bar v_P)\ \AS\ g_P),
}
\qdef{
q_{P,i}= & \CONSTRUCT & \grp{g_P\ \texttt{a c:Plus ; c:in}\ g_{P_i}}\\
 & \WHERE & \texttt{\{}\,\rwp{P_i}{g_{P_i}}\\
 & & \quad\BIND(\texttt{KEY\_PLUS}(\texttt{U},\theta(P),\bar v_P)\ \AS\ g_P)\,\texttt{\}},
}
\qrows{
\rewr(P)=\{q_{P,1},\ldots,q_{P,n}\}\cup\rewr(P_1)\cup\cdots\cup\rewr(P_n).
}
\item If $P=\opproj_{Y}(P')$, then, for a fresh variable $g_{P'}$ and the tuple $\bar Y$ of the variables in $Y$ in lexicographic order,
\qrows{
\rwp{P}{g_P}=\grp{\SELECT\ \bar Y\ \WHERE\ \grp{\rwp{P'}{g_{P'}}}}\\
\qquad\BIND(\texttt{KEY\_PLUS}(\texttt{P},\theta(P),\bar Y)\ \AS\ g_P),
}
\qdef{
q_P= & \CONSTRUCT & \grp{g_P\ \texttt{a c:Plus ; c:in}\ g_{P'}}\\
 & \WHERE & \texttt{\{}\,\rwp{P'}{g_{P'}}\\
 & & \quad\BIND(\texttt{KEY\_PLUS}(\texttt{P},\theta(P),\bar Y)\ \AS\ g_P)\,\texttt{\}},
}
\qrows{
\rewr(P)=\{q_P\}\cup\rewr(P').
}
\item If $P=\opfilter_\varphi(P')$, then $g_{P'}=g_P$, $\rwp{P}{g_P}=\rwp{P'}{g_{P'}}\ \FILTER(\varphi)$ and $\rewr(P)=\rewr(P')$.
\item If $P=\opdiff(P_1,P_2)$, then, for fresh variables $g_{P_1}$, $g_{P_2}$ and $s_P$ and the tuple $\bar v_{P_1}$ of the variables in scope at $P_1$ in lexicographic order,
\qrows{
\rwp{P}{g_P}=\rwp{P_1}{g_{P_1}}\\
\qquad\BIND(\texttt{KEY\_MINUS}(\theta(P),\bar v_{P_1})\ \AS\ g_P),
}
\qdef{
q^-_P= & \CONSTRUCT & \grp{s_P\ \texttt{a c:Plus ; c:in}\ g_{P_2}}\\
 & \WHERE & \texttt{\{}\texttt{\{}\,\rwp{P_1}{g_{P_1}}\\
 & & \quad\BIND(\texttt{KEY\_PLUS}(\texttt{S},\theta(P),\bar v_{P_1})\ \AS\ s_P)\,\texttt{\}}\\
 & & \phantom{\texttt{\{}}\grp{\rwp{P_2}{g_{P_2}}}\,\texttt{\}},\\[2pt]
q^+_P= & \CONSTRUCT & \texttt{\{}\,g_P\ \texttt{a c:Minus ; c:minuend}\ g_{P_1}\ \texttt{;}\\
 & & \quad\texttt{c:subtrahend}\ s_P\ \texttt{.}\ s_P\ \texttt{a c:Plus}\,\texttt{\}}\\
 & \WHERE & \texttt{\{}\,\rwp{P}{g_P}\\
 & & \quad\BIND(\texttt{KEY\_PLUS}(\texttt{S},\theta(P),\bar v_{P_1})\ \AS\ s_P)\,\texttt{\}},
}
\qrows{
\rewr(P)=\{q^-_P,q^+_P\}\cup\rewr(P_1)\cup\rewr(P_2).
}
\end{enumerate}
For a query $Q=\opproj_Y(P')$, item~(4) applies with $P=Q$ and the letter \texttt{A} in place of the letter \texttt{P}. For every row of $q_Q$ with restriction $\mu$ to $Y$, the gate $g_Q$ is the root of $\mu$, and the template of $q_Q$ additionally contains the \texttt{c:answerRoot} and \texttt{c:bind:} triples of~\eqref{eq:plan}.
\end{definition}

We highlight the difference $\opdiff(P_1,P_2)$ of item~(6). An \OPT{} without a filter abbreviates a union of a join and a difference (Section~\ref{sec:prelim}), and the identities of Appendix~\ref{app:proofs} reduce \MINUS{} to differences. For a mapping $\mu$ of $P_1$, the difference negates the disjunction over all mappings of $P_2$ that are compatible with $\mu$. The negation does not distribute over the disjunction and must therefore cover all rows of the join of $P_1$ and $P_2$ that extend $\mu$. SPARQLprov~\cite{hernandez2021sparqlprov} combines the rows of $\mu$ in a post-processing step that decodes the results, and NPCS~\cite{asma2024npcs} groups them by $\mu$ inside the endpoint. \sys instead splits the difference into two \CONSTRUCT{} queries. The subtrahend query $q^-_P$ names the sum gate $s_P$ after $\mu$ alone and binds $s_P$ in every row before the join with $P_2$, so the edges of all rows of $\mu$ merge at $s_P$. The minuend query $q^+_P$ evaluates only $\rwp{P_1}{g_{P_1}}$ and makes $s_P$ the subtrahend of the difference gate $g_P$.

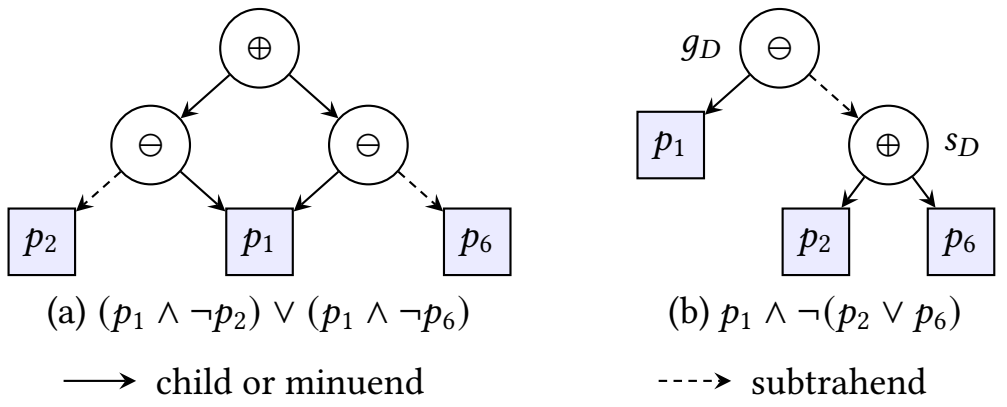
\begin{figure}[htbp]
\centering
\begin{tikzpicture}[>=stealth,
  gate/.style={draw, circle, minimum size=6.5mm, inner sep=0pt, font=\normalsize},
  leaf/.style={draw, fill=blue!8, minimum size=5.5mm, inner sep=1pt, font=\normalsize\bfseries},
  sub/.style={->, dash pattern=on 1.6pt off 1.2pt},
  scale=0.8, transform shape]
  \node[gate] (a0) at (1.5,1.6) {$\oplus$};
  \node[gate] (a1) at (0.6,0.8) {$\ominus$};
  \node[gate] (a2) at (2.4,0.8) {$\ominus$};
  \node[leaf] (ap2) at (-0.3,0) {$p_2$};
  \node[leaf] (ap1) at (1.5,0) {$p_1$};
  \node[leaf] (ap6) at (3.3,0) {$p_6$};
  \draw[->] (a0) -- (a1); \draw[->] (a0) -- (a2);
  \draw[->] (a1) -- (ap1); \draw[sub] (a1) -- (ap2);
  \draw[->] (a2) -- (ap1); \draw[sub] (a2) -- (ap6);
  \node[font=\normalsize] at (1.5,-0.6) {(a) $(p_1\land\neg p_2)\lor(p_1\land\neg p_6)$};
  \node[gate] (b0) at (5.8,1.6) {$\ominus$};
  \node[font=\normalsize, anchor=east] at (5.45,1.6) {$g_D$};
  \node[leaf] (bp1) at (4.9,0.8) {$p_1$};
  \node[gate] (bs) at (6.7,0.8) {$\oplus$};
  \node[font=\normalsize, anchor=west] at (7.05,0.8) {$s_D$};
  \node[leaf] (bp2) at (6.1,0) {$p_2$};
  \node[leaf] (bp6) at (7.3,0) {$p_6$};
  \draw[->] (b0) -- (bp1); \draw[sub] (b0) -- (bs);
  \draw[->] (bs) -- (bp2); \draw[->] (bs) -- (bp6);
  \node[font=\normalsize] at (6.1,-0.6) {(b) $p_1\land\neg(p_2\lor p_6)$};
  \draw[->] (-0.12,-1.15) -- (0.48,-1.15);
  \node[font=\normalsize, anchor=west] at (0.53,-1.15) {child or minuend};
  \draw[sub] (4.8,-1.15) -- (5.4,-1.15);
  \node[font=\normalsize, anchor=west] at (5.45,-1.15) {subtrahend};
\end{tikzpicture}
\caption{Circuits for Warfarin in Example~\ref{ex:diff}: (a) one $\ominus$-gate per row of the join, (b) the two queries of item~(6).}
\Description{Two small circuits. In (a), a sum gate has two difference gates as children, each with minuend p1 and one subtrahend leaf, p2 or p6. In (b), one difference gate has minuend p1 and as subtrahend a sum gate over p2 and p6. A legend below the circuits explains the solid edges to children or minuends and the dashed edges to subtrahends.}
\label{fig:diff}
\end{figure}

\begin{example}[Difference]\label{ex:diff}
The query
{\begin{verbatim}
  SELECT ?x WHERE { :Aspirin :interactsWith ?x
            MINUS { ?x :interactsWith ?y } }
\end{verbatim}}
\noindent returns the drugs with which Aspirin interacts and which interact with no drug. Since both operands of \MINUS{} bind $?x$, the \MINUS{} pattern equals the difference $D=\opdiff(\tp_1,\tp_4)$ for $\tp_4=(?x,\mathrm{interactsWith},?y)$, in which Warfarin has the Boolean provenance $p_1\land\neg(p_2\lor p_6)$. The join of $\tp_1$ and $\tp_4$ has the rows $(p_1,p_2)$ and $(p_1,p_6)$ for Warfarin. A single query that emits one $\ominus$-gate per row of the join yields the circuit of Figure~\ref{fig:diff}(a), which denotes $(p_1\land\neg p_2)\lor(p_1\land\neg p_6)$ and is therefore true in a world with the triples of $p_1$ and $p_2$ but without the triple of $p_6$, although $D$ removes Warfarin in this world. Under item~(6), the subtrahend query collects both rows in $s_D=\oplus(p_2,p_6)$, and the minuend query emits $g_D=\ominus(\{p_1\},s_D)$, so Figure~\ref{fig:diff}(b) denotes $p_1\land\neg(p_2\lor p_6)$.
\end{example}

The plan $\rewr(Q)$ contains one query for every join and every projection, one for every branch of a union and two for every difference. Every query reads only $G_{\mathrm{reif}}$ at the endpoint, so the queries are independent and can run concurrently.

\subsection{Factored basic graph patterns}\label{sec:factored}
We call the plan $\rewr(Q)$ of Definition~\ref{def:gamma} the \emph{flat plan}. It creates one product gate per derivation and therefore repeats every part that derivations share. A basic graph pattern with $m$ triple patterns has up to $|G|^m$ derivations. In Figure~\ref{fig:overview}, the product gates $\otimes(p_1,p_2,p_3)$ and $\otimes(p_4,p_5,p_3)$ of Omeprazole both contain $p_3$. For a query that projects a basic graph pattern, the \emph{factored plan} instead builds the circuit by variable elimination~\cite{dechter1999bucket} so that derivations share the gates of their common parts.

\begin{definition}[Factored plan]\label{def:factored}
Let $Q=\opproj_{Y}(\opjoin(\tp_1,\ldots,\tp_m))$, and let $Z$ contain the variables of $Q$ outside $Y$. A factor with scope $S$ maps some mappings of $S$ to gates, and its value at a mapping of a superset of $S$ is its value at the restriction to $S$. The base factor $F_j$ has the set $\var(\tp_j)$ of the variables of $\tp_j$ as its scope and maps every mapping $\rho$ with $\rho(\tp_j)\in G$ to the leaf $x_{\rho(\tp_j)}$. Eliminating a variable $z\in Z$ replaces the set $\mathcal A_z$ of the factors whose scopes contain~$z$ by a factor $H_z$ whose scope $U_z$ contains their other variables. For a mapping $\nu$ of $U_z$, let $A_z(\nu)$ contain the values $a$ such that every factor of $\mathcal A_z$ is defined at $\nu\cup\{z\mapsto a\}$. If $A_z(\nu)\neq\varnothing$, then
\[
 H_z(\nu)=\textstyle\bigoplus_{a\in A_z(\nu)}\,\bigotimes_{F\in\mathcal A_z}F(\nu\cup\{z\mapsto a\}).
\]
The factored plan eliminates the variables of $Z$ in some order. The root of an answer $\mu$ is the sum gate over the product of the remaining factors at $\mu$. The gate $H_z(\nu)$ is named by a tag of $Q$ and $z$ and the values of $\nu$, and the other gates as in Definitions~\ref{def:ids} and~\ref{def:gamma}.
\end{definition}

For the running query, $Z=\{?x,?y\}$. Eliminating $?x$ first multiplies $F_1$ and $F_2$, so $H_x$ has the scope $\{?y\}$ and maps Metformin to $h_x=\oplus(\otimes(p_1,p_2),\otimes(p_4,p_5))$ and Lisinopril to $h'_x=\oplus(\otimes(p_1,p_6))$. Eliminating $?y$ then multiplies $H_x$ and $F_3$, so $H_y$ has the scope $\{?z\}$ and maps Omeprazole to $\oplus(\otimes(h_x,p_3))$ and Clopidogrel to $\oplus(\otimes(h'_x,p_7))$. The induced width of an elimination order is the largest $|U_z|$, so the order $(?x,?y)$ has induced width~$1$. The root of Omeprazole denotes $p_3\land((p_1\land p_2)\lor(p_4\land p_5))$.

\begin{lemma}\label{lem:factored}
Assume that distinct names have distinct IRIs.\footnote{Distinct names share an IRI only if their 128-bit hashes collide. For $N$ gates, the probability of a collision is below $N^2/2^{128}$ in the random-oracle model (Appendix~\ref{app:builder}).} For every elimination order and every mapping $\mu$, the roots of $\mu$ in the factored plan and in the flat plan denote the same Boolean function.
\end{lemma}

With an elimination order of induced width~$w$, the factored plan creates $O(m\,|G|^{w+1})$ gates below the roots for fixed~$w$~\cite{dechter1999bucket}, whereas the flat plan creates up to $|G|^m$ product gates. For example, if Aspirin reaches Metformin through $k$ drugs and Metformin interacts with $l$ drugs, the flat plan creates $k\cdot l$ product gates for these walks and the factored plan $k+l$ (Appendix~\ref{app:full-hub}).

\sys writes the rows of every factor into a private namespace with SPARQL~1.1 Update, and every elimination reads the factors that earlier queries wrote there. The eliminations therefore run in sequence and need a writable endpoint.

\subsection{Transitive closures}\label{sec:paths}
For a closure-free $e$, an $e$-step is a pair $(u,v)\in\relpath{e}{G}$. The joins of Definition~\ref{def:gamma} match a fixed number of triples, whereas a closure atom matches walks of $e$-steps of unbounded length. As in the factored plan, \sys applies variable elimination along the walks, eliminates one node per step and builds one level per walk length. The levels form a variant of the Bellman--Ford circuit of Fan et al.~\cite{fan2025circuits}. For nodes $u$ and $v$, a gate $b^e_{uv}$ built by recursion on $e$ satisfies $\val{W}(\dec{b^e_{uv}})=\top$ exactly when $(u,v)\in\relpath{e}{W}$ (Appendix~\ref{app:full-paths}). For a source $s$, let $V_s$ contain $s$ and every node that $e$-steps reach from $s$, let $E_s$ contain the $e$-steps between nodes of $V_s$, and let $n=|V_s|$. Level~$1$ maps every node~$v$ to a gate $r_1(v)$ for the walks of length~$1$ from $s$ to~$v$, and level~$i+1$ sums out the last node~$u$ of the walks of level~$i$ from the product of $r_i$ and $b^e$:
\begin{align}
 r_1(v)&=\oplus\bigl(\{b^e_{sv}\mid(s,v)\in E_s\}\bigr),\label{eq:path-base}\\
 r_{i+1}(v)&=\oplus\bigl(\{\otimes(\{r_i(u),b^e_{uv}\})\mid(u,v)\in E_s,\ r_i(u)\ne g_\bot\}\bigr).\label{eq:path-step}
\end{align}
The closure gate of $v$ is
\begin{equation}
 R(v)=\oplus\bigl(\{r_i(v)\mid 1\le i\le \lambda\}\bigr),
 \label{eq:path-root}
\end{equation}
where $\lambda$ is the last level. The construction stops at the first empty level and runs at most $n$ levels, because a longer walk repeats a node and is absorbed by the walk without the repeated segment.

\begin{figure}[htbp]
\centering
\begin{tikzpicture}[>=stealth,
  ans/.style={draw, rounded corners, fill=gray!12, inner sep=2pt, minimum height=5mm, font=\normalsize\bfseries},
  lvl/.style={draw, circle, minimum size=6.5mm, inner sep=0pt, font=\normalsize\bfseries},
  shlvl/.style={draw, circle, fill=orange!18, minimum size=6.5mm, inner sep=0pt, font=\normalsize\bfseries},
  lab/.style={font=\normalsize, inner sep=1pt},
  scale=0.8, transform shape]
  \node[ans] (s) at (0,0.6) {Aspirin};
  \node[shlvl] (w1) at (1.35,1.2) {W};
  \node[lvl] (i1) at (1.35,0) {I};
  \node[lvl] (l2) at (2.6,1.2) {L};
  \node[lvl] (m2) at (2.6,0) {M};
  \node[lvl] (c3) at (3.85,1.2) {C};
  \node[lvl] (o3) at (3.85,0) {O};
  \node[lvl] (a4) at (5.1,1.2) {A};
  \node[shlvl] (w5) at (6.35,1.2) {W};
  \node[lvl] (i5) at (6.35,0) {I};
  \node[lvl] (l6) at (7.6,1.2) {L};
  \node[lvl] (m6) at (7.6,0) {M};
  \node[lvl] (c7) at (8.85,1.2) {C};
  \node[lvl] (o7) at (8.85,0) {O};
  \draw[->] (s) -- node[lab, above left] {$p_1$} (w1);
  \draw[->] (s) -- node[lab, below left] {$p_4$} (i1);
  \foreach \w/\i/\l/\m in {w1/i1/l2/m2, w5/i5/l6/m6} {
    \draw[->] (\w) -- node[lab, above] {$p_6$} (\l);
    \draw[->] (\w) -- node[lab, pos=0.4, right] {$p_2$} (\m);
    \draw[->] (\i) -- node[lab, below] {$p_5$} (\m);
  }
  \draw[->] (l2) -- node[lab, above] {$p_7$} (c3);
  \draw[->] (m2) -- node[lab, below] {$p_3$} (o3);
  \draw[->] (l6) -- node[lab, above] {$p_7$} (c7);
  \draw[->] (m6) -- node[lab, below] {$p_3$} (o7);
  \draw[->] (c3) -- node[lab, above] {$p_8$} (a4);
  \draw[->] (a4) -- node[lab, above] {$p_1$} (w5);
  \draw[->] (a4) -- node[lab, pos=0.4, right] {$p_4$} (i5);
  \foreach \x/\i in {1.35/1, 2.6/2, 3.85/3, 5.1/4, 6.35/5, 7.6/6, 8.85/7} \node[font=\normalsize] at (\x,-0.65) {\i};
\end{tikzpicture}
\caption{Levels 1 to 7 of \texttt{:Aspirin :interactsWith+ ?y}.}
\Description{A layered graph. The source Aspirin reaches Warfarin and Ibuprofen at level 1, Lisinopril and Metformin at level 2, Clopidogrel and Omeprazole at level 3 and Aspirin at level 4, and the same pattern repeats at levels 5 to 7. The edges are labeled with the triples p1 to p8, and the two Warfarin circles at levels 1 and 5 are orange.}
\label{fig:levels}
\end{figure}
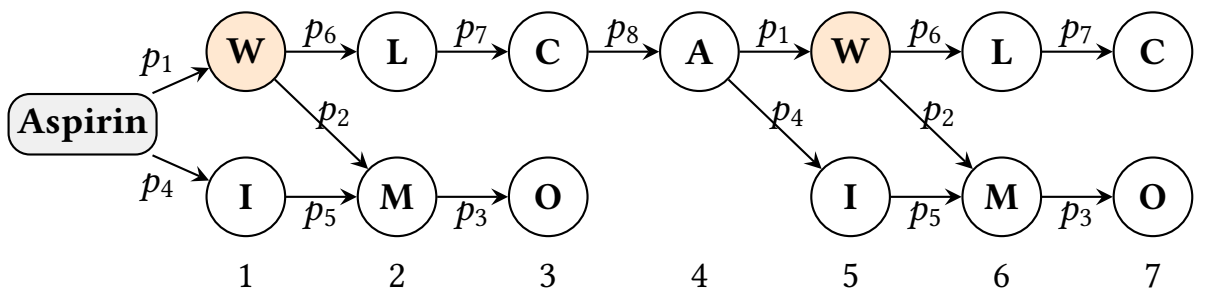

For the query \texttt{SELECT ?y WHERE \{ :Aspirin :interactsWith+ ?y \}} over the graph of Figure~\ref{fig:running}, $V_s$ contains the seven drugs, and Figure~\ref{fig:levels} shows the levels, where a circle is the sum gate of a drug at one level, labeled by the initial of the drug, and an arrow multiplies by the gate of one step. At level~$2$, the sum gate of Metformin adds the product of the level-$1$ gate of Warfarin with $p_2$ and the product of the level-$1$ gate of Ibuprofen with $p_5$, and denotes $(p_1\land p_2)\lor(p_4\land p_5)$. The cycle through Warfarin, Lisinopril and Clopidogrel returns to Aspirin at level~$4$ and keeps every level nonempty, so the construction runs $n=7$ levels. The closure gate of Warfarin sums the levels~$1$ and~$5$, and $p_1$ absorbs the level-$5$ product $p_1\land p_6\land p_7\land p_8$ for the walk around the cycle.

For a source-bound closure atom~$C$ (Section~\ref{sec:prelim}), the input relation $\mathcal I_C$ contains the mappings of the rows of the operands before $C$. The set $S_C$ contains the sources that $\mathcal I_C$ binds, or only $\alpha$ when the source is a constant, and the plan runs the construction once for every source $s\in S_C$. For a target $v$ returned from $s$, the mapping $\mu^C_{s,v}$ has the gate $g_C(s,v)=R(v)$. For $\diamond=*$, $g_C(s,s)$ is instead the zero-length gate of $s$, which is $g_\top$ for a constant source and the sum gate $o_s$ over the leaves of the triples of $G$ with subject or object $s$ for a variable source. The plan materializes the mappings and their gates as a relation, and $\rwp{C}{g_C}$ reads the relation in place of the reification of a triple pattern. In the pattern \texttt{\{ :Aspirin :interactsWith ?x . ?x :interactsWith+ ?y \}}, the triple pattern binds the source $?x$, so $S_C=\{\mathrm{Warfarin},\mathrm{Ibuprofen}\}$, and the row with $?x=\mathrm{Warfarin}$ and $?y=\mathrm{Lisinopril}$ has the product gate $\otimes(p_1,g_C(\mathrm{Warfarin},\mathrm{Lisinopril}))$. A path pattern with a constant target reduces to a closure atom (Appendix~\ref{app:proofs}).

\begin{theorem}\label{thm:path}
Assume that distinct names have distinct IRIs. Let $C$ be a closure atom in a query of \frag, let $s\in S_C$, and let $v$ be a target that the construction returns from $s$, with $v=s$ when $\alpha$ is $?y$. For every possible world $W\subseteq G$, $\val{W}(\dec{g_C(s,v)})=\top$ exactly when $C$ returns $\mu^C_{s,v}$ on $W$, so $g_C(s,v)$ denotes $\BProv{C}{\mu^C_{s,v}}$.
\end{theorem}

For every source, the levels form an acyclic circuit with $O(n(n+|E_s|))$ gates. \sys emits the gates $b^e_{uv}$ and level~$1$ with \CONSTRUCT{} queries and every further level with one more round. By default, every round writes its level into a private namespace with SPARQL~1.1 Update, where the next round reads it. If $C$ forms the whole query, every round can instead send the previous level in a \VALUES{} clause, so that the construction runs on a read-only endpoint. If $C$ is a proper subpattern of the query, the materialization of its relation requires a writable endpoint.

\subsection{Correctness and probabilities}\label{sec:correctness}
{\setlength{\parindent}{0pt}\begin{definition}[Soundness and completeness]\label{def:sound}
A plan for a query $Q$ is \emph{sound} if every mapping $\mu$ with a root in its circuit over $G_{\mathrm{reif}}$ satisfies
\begin{equation}
 \dec{\mathrm{root}(\mu)}=\BProv{Q}{\mu},
 \label{eq:query-correct}
\end{equation}
and \emph{complete} if this circuit has a root for every answer of $Q$.
\end{definition}}

In the circuit of a sound and complete plan, the weighted model count of the root of every answer is the marginal probability of the answer, and every mapping without a root has probability~$0$.

The plan evaluates its queries once over $G_{\mathrm{reif}}$, and every world is a subset of $G$. By induction on the subpatterns of $Q$, every answer $\mu$ of $Q$ is therefore the restriction of a row of $\rwp{Q}{g_Q}$ and receives a root. Conversely, other rows can restrict to mappings that $Q$ returns in no world, because item~(6) of Definition~\ref{def:gamma} keeps every row of the minuend in the rewritten pattern of a difference. For $\opdiff(P,P)$, every mapping $\mu$ of $P$ is compatible with itself, and the difference gate of $\mu$ denotes $\bot$. A mapping that is not an answer can therefore receive a root, and soundness requires this root to denote $\bot$.

Section~\ref{sec:prelim} defines the function of a gate by recursion over the child relation, so this relation must be acyclic. The gates of a triple pattern are the leaves that its matches bind, and the gates of any other subpattern $P$ are the gates that its queries emit. Every child of a gate of $P$ is a gate of a proper subpattern of $P$, except that the child $s_P$ of a difference gate $g_P$ is a gate of $P$ itself and has children only among the gates of $P_2$. Within a closure atom, every edge leads to a lower level, and every factor of the factored plan is built from earlier factors. The child relation is therefore acyclic.

\begin{lemma}\label{lem:names}
Assume that distinct names have distinct IRIs. Let $Q$ be a query of \frag, let $P$ be a union, a projection or a difference in $Q$, and let $\mu$ be a mapping whose domain consists of variables in scope at $P$. In the circuit of $\rewr(Q)$ over $G_{\mathrm{reif}}$, the gate $g_P$ named after $P$ and $\mu$ denotes $\BProv{P}{\mu}$. If $P=\opdiff(P_1,P_2)$, the gate $s_P$ named after $P$ and $\mu$ denotes $\bigvee_{\nu\sim\mu}\BProv{P_2}{\nu}$.
\end{lemma}

The root of a mapping $\mu$ is the gate $g_Q$ named after $Q$ and $\mu$, so Lemma~\ref{lem:names} yields Equation~\eqref{eq:query-correct}, and Lemma~\ref{lem:factored} carries Equation~\eqref{eq:query-correct} over to the factored plan.

\begin{theorem}\label{thm:query}
Assume that distinct names have distinct IRIs. For every query $Q=\opproj_{Y}(P)$ of \frag, the plan $\rewr(Q)$ is sound and complete, and so is the factored plan when $P$ is a basic graph pattern.
\end{theorem}

The theorem and the computation of $\Pr(\mu)$ extend to a triple with several identifiers (Appendix~\ref{app:builder}).

Finally, \sys compiles every root into an OBDD with CUDD~\cite{somenzi2015cudd} or into a d-DNNF with d4~\cite{lagniez2017d4} and computes the weighted model count of the result (Appendix~\ref{app:full-impl}).

\section{Evaluation}\label{sec:eval}
In this section, we measure the overhead of circuit construction (Section~\ref{sec:eval:overhead}), the growth of the endpoint time and of the provenance size with the number of derivations per answer (Section~\ref{sec:eval:sharing}) and the end-to-end time of PQE (Section~\ref{sec:eval:wikidata}).

\noindent\textbf{Data.}
WatDiv~\cite{aluc2014watdiv} generates graphs of 10.9 and 109 million triples and queries from 20 templates of basic graph patterns, to which we add five \OPT{} and five \MINUS{} templates. We report the WatDiv times under RDF-star annotations. gMark~\cite{gmark} generates chain queries with $k$ triple patterns and branching factor $W$, the number of choices at every step, over graphs of 500{,}000 triples. On the 1.78 billion statements of Wikidata~\cite{vrandecic2014wikidata}, each identified by its statement node, we run the 136 WDBench queries~\cite{angles2022wdbench} that NPCS publishes and 45 property-path queries from WDBench. All probabilities are drawn uniformly from $(0,1)$.

\noindent\textbf{Engines.}
GraphDB 10.7.6~\cite{graphdb}, Fuseki 5.4.0 with TDB2~\cite{fuseki} and Oxigraph 0.5.9~\cite{oxigraph} run locally without extensions and receive SPARQL~1.1 requests and, for RDF-star, SPARQL-star requests.

\noindent\textbf{Baselines.}
The plain query (Base) runs on $G$, and the reified query (Reification) runs on $G_{\mathrm{reif}}$ without provenance. We compare with NPCS~\cite{asma2024npcs}, the state-of-the-art SPARQL provenance rewriting. For PQE, we merge the identical subexpressions of all answers into one DAG and compile it with CUDD~3.0.0. We run our reimplementation of the rewriting rules of NPCS, since its published JAR deviates from these rules beyond basic graph patterns (Appendix~\ref{app:artifacts}).

\noindent\textbf{Protocol.}
The endpoint time of a method ends when the response or the circuit is stored, and the end-to-end time ends when all probabilities are written. We report the median of five runs after a warm-up run, medians over sets of queries and medians of paired ratios. The time budget is the same for every method and ranges from 600 to 5000\,s (Appendix~\ref{app:experiments}).

\subsection{Overhead of circuit construction}\label{sec:eval:overhead}
\begin{figure}[htbp]
\centering
\vspace{-0.1cm}
\includegraphics[width=0.95\columnwidth]{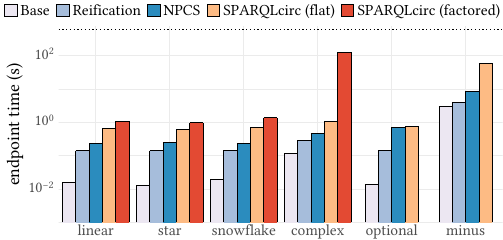}
\vspace{-0.1cm}
\caption{Endpoint time on WatDiv (GraphDB, 10.9M triples).}
\Description{A logarithmic bar chart gives, for the linear, star, snowflake and complex templates of WatDiv and for the templates with OPTIONAL and MINUS, the endpoint time in seconds of Base, Reification, NPCS and the flat plan of SPARQLcirc on GraphDB with 10.9 million triples, together with the factored plan on the templates of basic graph patterns. A dotted line marks the budget. On the linear, star and snowflake templates, Base takes about 15 ms, NPCS about 250 ms and the flat plan about 700 ms.}
\vspace{-0.2cm}
\label{fig:watdiv}
\end{figure}

Following the evaluation of NPCS~\cite{asma2024npcs}, we measure the overhead of provenance on WatDiv relative to the plain query and the reified query. With 10.9 million triples on GraphDB, the reified query takes 8.7 times as long as the plain query, NPCS 1.9 times as long as the reified query and the flat plan 2.6 times as long as NPCS (Figure~\ref{fig:watdiv}), or 0.87\,s against 0.46\,s end to end. The factored plan takes 4.3 times as long as NPCS. On the templates with \MINUS{}, the flat plan takes 3.4 to 7.8 times as long as NPCS across the engines and sizes and exceeds the budget more often.  Appendix~\ref{app:exp:watdiv} gives all 30 templates on all three engines, sizes and encodings.

One circuit saves time only when the answers share subexpressions. On WatDiv, merging the identical subexpressions of NPCS does not reduce the median size of its provenance, so the flat plan pays only the overhead of a \CONSTRUCT{} query per operator and of storing the returned triples, whereas NPCS sends one \SELECT{} query. Most of this overhead is fixed per query, because with ten times as many triples, the plain query takes 3.9 times as long and the flat plan 1.5 times as long. The factored plan also writes every factor and runs the eliminations in sequence, and for \MINUS{}, the subtrahend query evaluates the join of both operands.

\subsection{\texorpdfstring{Scalability with the number of derivations}{Scalability with the number of derivations}}\label{sec:eval:sharing}
\begin{figure}[htbp]
\centering
\includegraphics[scale=0.8]{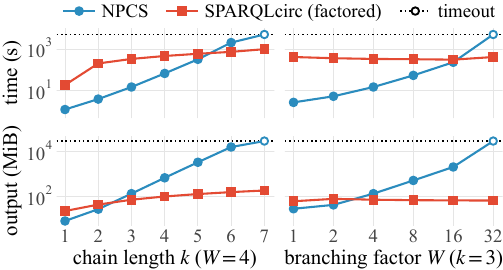}
\vspace{-0.2cm}
\caption{Endpoint time and output size on gMark (GraphDB).}
\Description{Four logarithmic line charts in two rows. The top row plots the endpoint time of NPCS and of the factored plan against the chain length k at W equal to 4 and against the number W of choices per step at k equal to 3. The bottom row plots the size of the response of NPCS and of the circuit of the factored plan for the same two sweeps.}
\label{fig:gmark}
\end{figure}

We control the number of derivations per answer through the chain length $k$ and the branching factor $W$ of gMark and compare the endpoint time and the output size of NPCS and the factored plan (Figure~\ref{fig:gmark}). At $W=4$, the response of NPCS grows about fivefold per triple pattern, and the circuit grows linearly. At $k=6$, the response reaches 15.9\,GiB, 101 times the size of the circuit. NPCS is faster for $k\le5$ at $W=4$ and for $W\le16$ at $k=3$, and the factored plan is faster for larger values, by a factor of about three at $k=6$. At $k=7$ and at $W=32$, NPCS exceeds the budget of 5000\,s, whereas the factored plan completes.

NPCS lists every derivation of an answer, so its response grows with the number of derivations. The factored plan shares the common parts of the derivations, so its circuit grows only with the number of triple patterns. To share gates, the factored plan writes a factor from all matches of every triple pattern to the endpoint and takes 313 to 418\,s in total at $k=3$. This time hardly depends on $W$, because with a fixed number of triples, a larger $W$ yields fewer nodes, and each elimination creates at most one sum gate per node.

\subsection{\texorpdfstring{End-to-end PQE on Wikidata}{End-to-end PQE on Wikidata}}\label{sec:eval:wikidata}
We measure the end-to-end time of PQE on the WDBench queries and the property-path queries of Wikidata. We report the flat plan, which completes more queries than the factored plan (Appendix~\ref{app:exp:wikidata}). On the WDBench queries, the flat plan is faster than NPCS on 40 of the 82 queries that both methods complete on GraphDB and on 41 of 74 on Oxigraph, depending on the total number of derivations of the answers of a query. For each range of this number, Table~\ref{tab:wikidata} gives the number of faster queries per method and the median ratio of the time of the flat plan to that of NPCS. NPCS is faster on every query with fewer than 100 derivations, and the flat plan is faster on most queries with at least 10{,}000 derivations, where it takes 0.57 times as long as NPCS at the median. NPCS does not support transitive closures, whereas \sys completes 35 of the 45 property-path queries on both engines, in a median of 3.4\,s on GraphDB. The largest of these circuits has 1.2 million gates for 38{,}757 answers.

\begin{table}[t]
\caption{End-to-end comparison on Wikidata.}
\label{tab:wikidata}
\small
\centering
\begin{tabular*}{0.9\columnwidth}{@{}l@{\extracolsep{\fill}}rrrrrr@{}}
\toprule
 & \multicolumn{3}{c}{GraphDB} & \multicolumn{3}{c}{Oxigraph}\\
\cmidrule(lr){2-4}\cmidrule(l){5-7}
Derivations & Flat & NPCS & Ratio & Flat & NPCS & Ratio\\
\midrule
Fewer than $10^2$ & 0 & 17 & 1.20 & 0 & 17 & 1.11\\
$10^2$ to $10^4$ & 12 & 21 & 1.06 & 18 & 14 & 0.99\\
At least $10^4$ & 28 & 4 & 0.57 & 23 & 2 & 0.58\\
\midrule
All & 40 & 42 & 1.01 & 41 & 33 & 0.98\\
\bottomrule
\end{tabular*}
\end{table}

The advantage of the flat plan on queries with many derivations stems from the preprocessing that NPCS requires before compilation. NPCS parses its expressions, which grow with the number of derivations, and merges them into one DAG. On the queries with at least 10{,}000 derivations, this takes about five times as long as reading the circuit of the flat plan.

\vspace{-0.4cm}
\section{Related Work}\label{sec:related}

\noindent\textbf{Probabilistic relational databases.}
Exact PQE has PTIME versus \#P\nobreakdash-hard dichotomies for unions of conjunctive queries~\cite{dalvi2012dichotomy} and for two classes of queries with negation~\cite{fink2016negation}, where even a difference of tractable queries can be \#P\nobreakdash-hard~\cite{khanna2011difference}. On a standard database system, Gatterbauer and Suciu compute exact probabilities or upper bounds for conjunctive queries without self-joins with SQL queries that obtain products through sums of logarithms~\cite{gatterbauer2017dissociation}. SPARQL~1.1 offers no logarithm, exponential or product aggregate~\cite{w3c-sparql11}, so \sys compiles the circuit that the endpoint returns. SPROUT approximates, in a PostgreSQL extension, the probability of the lineage that rewritten SQL queries compute~\cite{fink2013sprout}, and ProvSQL builds provenance circuits as a PostgreSQL extension~\cite{senellart2018provsql}. Unlike ProvSQL, \sys computes gate names in SPARQL~1.1 queries.

\noindent\textbf{Probabilistic RDF.}
Earlier systems attach distributions to object values~\cite{udrea2006prdf} or to node labels~\cite{lian2011probrdf} and test triples or matches against a probability threshold. Huang and Liu compile the lineage of each answer of a conjunctive SPARQL query into a binary decision diagram and approximate the probability of transitive properties on acyclic graphs~\cite{huang2009probrdf}. HAPPI computes the probability of each answer of a conjunctive query from a symbolic provenance expression~\cite{gaur2021happi}. These systems use dedicated processors and treat every answer separately. \sys builds one circuit for all answers with SPARQL~1.1 queries and computes exact probabilities also for non-monotone queries and transitive closures on cyclic graphs.

\noindent\textbf{Provenance for SPARQL.}
Semiring provenance~\cite{green2007provenance} with monus~\cite{geerts2010krelations} captures \OPT{} and \MINUS{}~\cite{damasio2012provenance}, and spm-semirings in addition preserve the equivalences of SPARQL queries~\cite{geerts2016algebraic}. Dividino et al.\ annotate every SPARQL answer with a Boolean formula that also covers \OPT{} and difference through an extended SPARQL syntax~\cite{dividino2009metaknowledge}, and TripleProv computes provenance in a dedicated engine~\cite{wylot2014tripleprov}. Ganepola et al.\ compute the why-provenance of monotone Core-GQL queries by rewriting~\cite{ganepola2026whyprov}. SPARQLprov and NPCS return one provenance expression per answer~\cite{hernandez2021sparqlprov,asma2024npcs,asma2026npcs}, whereas \sys returns one circuit for all answers, and the gate names agree across queries and engines.

\noindent\textbf{Provenance under recursion.}
Deutch et al.\ build one layered Datalog provenance circuit for all derived tuples~\cite{deutch2014circuits}. Bellman--Ford-style circuits~\cite{fan2025circuits} and algorithms~\cite{ramusat2021graph} compute single-source provenance. ProvSQL~1.10 compiles recursive reachability queries into deterministic decomposable circuits~\cite{provsql2026recursion}. There is a regular path query whose provenance has no d\nobreakdash-DNNF or OBDD of size polynomial in the graph~\cite{amarilli2025approximating}. The circuit of a closure atom has polynomial size, but computing the probability of its root is \#P\nobreakdash-hard~\cite{valiant1979}. \sys builds a Bellman--Ford-style circuit level by level with SPARQL~1.1 requests and stops at the first empty level.
\section{Conclusion}\label{sec:concl}
We presented \sys, a rewriting method for exact PQE on SPARQL~1.1 endpoints. The plans name every gate by its content, so independent \CONSTRUCT{} queries build one provenance circuit for all answers. The plans are sound and complete for joins, unions, projection, filters, \OPT{}, difference and transitive closures. The ratio of the end-to-end time of PQE with one circuit to that with per-answer provenance expressions decreases with the number of derivations. Future work includes federated provenance circuits.

\begin{acks}
The project was funded by the Deutsche Forschungsgemeinschaft (DFG, German Research Foundation) - SFB~1574 - Project number 471687386 (``Circular Factory''). The authors thank the International Max Planck Research School for Intelligent Systems (IMPRS-IS) for supporting Jingcheng Wu.
\end{acks}

\AtBeginEnvironment{thebibliography}{\interlinepenalty=10000}
\bibliographystyle{ACM-Reference-Format}
\bibliography{references}

\appendix
\section{Proofs}\label{app:proofs}

\subsection{Factored basic graph patterns}\label{app:factored}
{\noproofindent\begin{proof}[Proof of Lemma~\ref{lem:factored}]
Let $V_Q$ be the variables of $Q$, let $\adom(G)$ contain the terms that occur in $G$, and let $(z_1,\ldots,z_k)$ be the elimination order. We identify a factor with the map that sends every mapping of the scope to the Boolean function of the gate of the factor at the mapping, and to $\bot$ where the factor has no gate. The base factor $F_j$ thus maps $\rho$ to $x_{\rho(\tp_j)}$ when $\rho(\tp_j)\in G$ and to $\bot$ otherwise. For a mapping $\mu$ with $\dom(\mu)\subseteq Y$, the root of $\mu$ in the flat plan denotes
\begin{equation}
 \Phi^{\mathrm{flat}}(\mu)=
 \bigvee_{\substack{{\rho}:V_Q\to\adom(G)\\{\rho}|_{Y}=\mu}}\ \bigwedge_{j=1}^{m}F_j\bigl(\rho|_{\var(\tp_j)}\bigr).
 \label{eq:flat-formula}
\end{equation}
Let $\mathcal F_i$ be the factors after eliminating $z_1,\ldots,z_i$. For a binding $\rho$ of the remaining variables, let
\begin{equation}
 I_i(\rho)=
 \bigvee_{\kappa:\{z_1,\ldots,z_i\}\to\adom(G)}
 \bigwedge_{f\in\mathcal F_0}
 f\bigl((\rho\cup\kappa)|_{\mathrm{scope}(f)}\bigr).
 \label{eq:elim-invariant}
\end{equation}
We prove that the conjunction of the factors in $\mathcal F_i$ at $\rho$ equals $I_i(\rho)$.
For $i=0$, Equation~\eqref{eq:elim-invariant} contains the unique empty assignment and equals the original factor conjunction.
Assume the claim for $i-1$, let $z=z_i$, and let $\rho_a=\rho\cup\{z\mapsto a\}$ for $a\in\adom(G)$. The scopes of the factors outside $\mathcal A_z$ do not contain $z$. The claim for $i-1$ and distributivity therefore give
\begin{align*}
 I_i(\rho)
 &=\bigvee_{a\in\adom(G)}
 \left[
 \bigwedge_{f\in\mathcal A_z} f(\rho_a|_{\mathrm{scope}(f)})
 \land
 \bigwedge_{f\notin\mathcal A_z} f(\rho|_{\mathrm{scope}(f)})
 \right]\\
 &=H_z(\rho|_{U_z})\land
 \bigwedge_{f\notin\mathcal A_z} f(\rho|_{\mathrm{scope}(f)}).
\end{align*}
The second equality holds because the terms of the values $a\notin A_z(\rho|_{U_z})$ are $\bot$, and the disjunction of the other terms is the function of the gate $H_z(\rho|_{U_z})$ of Definition~\ref{def:factored}, or $\bot$ when $A_z(\rho|_{U_z})$ is empty. The result is the conjunction after replacing $\mathcal A_z$ by $H_z$. At $i=k$, restriction to $\mu$ yields Equation~\eqref{eq:flat-formula}, so the roots of $\mu$ in both plans denote the same Boolean function.

\end{proof}}

\subsection{Transitive closures}\label{app:paths}
{\noproofindent\begin{proof}[Proof of Theorem~\ref{thm:path}]
The gates $b^e_{uv}$ form an acyclic circuit and satisfy the equivalence of Section~\ref{sec:paths} (Appendix~\ref{app:full-paths}). For an $e$-walk $\pi=(v_0,\ldots,v_\ell)$ in $E_s$, we write $\pi:v_0\leadsto v_\ell$ and $|\pi|=\ell$, and let
\[
 \psi(\pi)=\bigwedge_{j=1}^{\ell}\dec{b^e_{v_{j-1}v_j}}.
\]
We prove by induction on $i\ge1$ that
\begin{equation}
 \dec{r_i(v)}=
 \bigvee_{\substack{\pi:s\leadsto v\\|\pi|=i}}\psi(\pi).
 \label{eq:path-level-invariant}
\end{equation}
For $i=1$, Equation~\eqref{eq:path-base} is the disjunction over all one-edge $e$-walks from $s$ to $v$. Assume Equation~\eqref{eq:path-level-invariant} for $i$. Equation~\eqref{eq:path-step} gives
\begin{align*}
 \dec{r_{i+1}(v)}
 &{}=\bigvee_{(u,v)\in E_s}
 \bigl(\dec{r_i(u)}\land\dec{b^e_{uv}}\bigr)\\
 &{}=\bigvee_{(u,v)\in E_s}
 \bigvee_{\substack{\pi':s\leadsto u\\|\pi'|=i}}
 \bigl(\psi(\pi')\land\dec{b^e_{uv}}\bigr)\\
 &{}=\bigvee_{\substack{\pi:s\leadsto v\\|\pi|=i+1}}\psi(\pi).
\end{align*}
Appending $(u,v)$ to every walk of length $i$ ending at $u$ gives exactly the walks of length $i+1$ ending at $v$.

If $v\ne s$, a shortest positive $e$-walk from $s$ to $v$ contains no repeated node and has length at most $n-1$, because removing the segment between equal nodes yields a shorter walk. If $v=s$, the same deletion argument makes a shortest positive closed walk a simple cycle of length at most $n$. The construction runs $\lambda\le n$ levels, and Equation~\eqref{eq:path-root} gives
\[
 \dec{R(v)}=\bigvee_{\substack{\pi:s\leadsto v\\1\le|\pi|\le \lambda}}\psi(\pi).
\]
If the construction stopped because level $\lambda+1$ is empty, no walk exceeds $\lambda$, since every longer walk has a prefix of length $\lambda+1$. Otherwise $\lambda=n$. The closure gate $R(v)$ is therefore true in $W$ whenever $(s,v)\in\relpath{e^+}{W}$, because a shortest such walk has length at most $\lambda$. Conversely, every satisfied term denotes an $e$-walk in $W$, so for every $W\subseteq G$
\begin{equation}
 \val{W}(\dec{R(v)})=\top
 \,\,\Longleftrightarrow\,\,
 (s,v)\in\relpath{e^+}{W}.
 \label{eq:path-correct}
\end{equation}
Let $\mu=\mu^C_{s,v}$. For $\diamond=+$, $g_C(s,v)=R(v)$, and $C$ returns $\mu$ on $W$ exactly when $(s,v)\in\relpath{e^+}{W}$. For $\diamond=*$ and $v\ne s$, $g_C(s,v)=R(v)$ as well, and the same equivalence holds because a walk from $s$ to $v\ne s$ has positive length. For $\diamond=*$ and $v=s$, $g_C(s,s)$ is the zero-length gate of $s$. For a constant source, the zero-length gate is $g_\top$, and $C$ returns $\mu$ on every $W$. For a variable source, the zero-length gate is $o_s$, which is true in $W$ exactly when $s$ is the subject or object of a triple of $W$. The atom $C$ returns $\mu$ on $W$ under the same condition, because the first $e$-step of a closed walk at $s$ uses a triple of $W$ whose subject or object is $s$. In every case, $\val{W}(\dec{g_C(s,v)})=\top$ exactly when $C$ returns $\mu$ on $W$, which holds exactly when $\val{W}(\BProv{C}{\mu})=\top$ by the definition of $\BProv{C}{\cdot}$ in Section~\ref{sec:prelim}. The two Boolean functions therefore agree under every valuation of $X$.

Each of the at most $n$ levels has at most $|V_s|$ sum gates and $|E_s|$ product gates, so the levels have $O(n(n+|E_s|))$ gates. One further sum gate per target forms $R(v)$. Level $i+1$ references only the gates $b^e_{uv}$ and level-$i$ gates, so the level order is topological.
\end{proof}}

\subsection{Correctness of the plans}
{\noproofindent\begin{proof}[Proof of Lemma~\ref{lem:names}]
Under the assumption, two gates have the same IRI exactly when they have the same name, and we use three facts. (1)~Two product gates have the same IRI if and only if their lists of children are equal up to order, because the name $(\texttt{T},\operatorname{sort}(c_1,\ldots,c_n))$ is the same for two lists of children exactly when the lists are equal up to order. (2)~Two sum gates, or two difference gates, have the same IRI if and only if they have the same letter, the same mapping and subpatterns equal up to the identities of Definition~\ref{def:ids}. The name $(L,\theta(P),\tau(a_1),\ldots,\tau(a_k))$ lists the values $a_j$ of the variables in scope at $P$ under the mapping, with $a_j=\mathsf{unbound}$ for an unbound variable, $\theta(P)$ determines $P$ up to these identities, which preserve Boolean provenance, and $\tau$ is injective. (3)~The IRI of a gate depends only on the row, because a key expression applies only functions whose values SPARQL~1.1 fixes on IRIs and literals (Appendix~\ref{sup:builder}), and $G_{\mathrm{reif}}$ contains no blank nodes.

For a pattern $P$ and a mapping $\mu$, let $\bar v_P$ be the tuple of the variables in scope at $P$ in lexicographic order, as in Definition~\ref{def:gamma}, let $\cand{P}{\mu}$ be the gates bound by the rows of $\rwp{P}{g_P}$ that restrict to $\mu$ on $\bar v_P$, and let
\begin{equation}
 \Phi_P(\mu)=\bigvee_{g\in\cand{P}{\mu}}\dec g,
 \qquad \bigvee\varnothing=\bot.
 \label{eq:candidate-formula}
\end{equation}
We then prove $\Phi_P(\mu)=\BProv{P}{\mu}$ by structural induction on $P$, for every subpattern $P$ that is not a closure atom with a variable source and for every mapping $\mu$. The base cases are triple patterns and closure atoms. For a compound pattern, the induction hypothesis is the claim for each of its direct subpatterns and every mapping. For a closure atom we prove the claim at the mappings whose source lies in $S_C$, and the join and difference cases use the source-bound condition to apply it.

\emph{Triple pattern.} For a matching triple $t$, $\cand{P}{\mu}=\{x_t\}$ and
$\Phi_P(\mu)=x_t=\BProv{P}{\mu}$. For every other mapping, both sides are $\bot$.

\emph{Closure atom.} Let $P=C=(\alpha,e^\diamond,?y)$. Every row of $C$ restricts on $\bar v_C$ to a mapping $\mu^C_{s,v}$ with $s\in S_C$ and a returned target $v$, and all rows for one pair $(s,v)$ bind the gate $g_C(s,v)$ of Section~\ref{sec:paths}. For $s\in S_C$, Theorem~\ref{thm:path} gives $\Phi_C(\mu^C_{s,v})=\dec{g_C(s,v)}=\BProv{C}{\mu^C_{s,v}}$. If the construction does not return $v$, then $(s,v)\notin\relpath{e^+}{W}$ for every $W\subseteq G$, so $\BProv{C}{\mu^C_{s,v}}=\bot$ and both sides are $\bot$. Thus
\begin{equation}
 \Phi_C(\mu)=\BProv{C}{\mu}\quad\text{for every $\mu$ whose source lies in $S_C$.}
 \label{eq:atom-relative}
\end{equation}
For a constant source, $S_C=\{\alpha\}$, so Equation~\eqref{eq:atom-relative} holds for every $\mu$.

\emph{Filter.} Let $P=\opfilter_\varphi(P')$. If $\mu$ satisfies $\varphi$, then $\cand{P}{\mu}=\cand{P'}{\mu}$ and the induction hypothesis gives $\Phi_P(\mu)=\BProv{P'}{\mu}=\BProv{P}{\mu}$. If $\mu$ does not satisfy $\varphi$, item~(5) keeps \FILTER($\varphi$) in the rewritten pattern, so $\cand{P}{\mu}=\varnothing$ and $\Phi_P(\mu)=\bot$, and $\BProv{P}{\mu}=\bot$ by the filter case of Section~\ref{sec:prelim}.

\emph{Join.} Let $P=\opjoin(P_1,\ldots,P_{n})$. The gates of $\cand{P}{\mu}$ are the product gates of the compatible decompositions $\mu=\mu_1\cup\cdots\cup\mu_{n}$ and choices $g_i\in\cand{P_i}{\mu_i}$. Fix a compatible decomposition $\mu=\mu_1\cup\cdots\cup\mu_{n}$ and a world $W\subseteq G$. We show that $\val{W}$ satisfies $\bigwedge_i\Phi_{P_i}(\mu_i)$ exactly when $\val{W}$ satisfies $\bigwedge_i\BProv{P_i}{\mu_i}$. For an operand that is not a closure atom with a variable source, the induction hypothesis gives $\Phi_{P_i}(\mu_i)=\BProv{P_i}{\mu_i}$. Let $P_i=C$ be a closure atom with a variable source. If $\val{W}$ satisfies every $\Phi_{P_j}(\mu_j)$, then $C$ has a row at $\mu_i$, so the source of $\mu_i$ lies in $S_C$ and Equation~\eqref{eq:atom-relative} applies. If $\val{W}$ satisfies every $\BProv{P_j}{\mu_j}$, induction on $j$ shows that $\val{W}$ satisfies $\Phi_{P_j}(\mu_j)$ for every $j<i$. Every such $P_j$ then has a row at $\mu_j$, so $\rho=\bigcup_{j<i}\mu_j$ belongs to $\mathcal I_C$. The source-bound condition makes $\rho$ bind the source, and compatibility gives $\rho(\alpha)=\mu_i(\alpha)$. The source of $\mu_i$ therefore lies in $S_C$, and Equation~\eqref{eq:atom-relative} applies. Distributivity over the choices of the gates $g_i$ and the equivalence just shown give
\[
 \Phi_P(\mu)
 =\bigvee_{\mu=\cup_i\mu_i}\bigwedge_{i=1}^{n}\Phi_{P_i}(\mu_i)
 =\bigvee_{\mu=\cup_i\mu_i}\bigwedge_{i=1}^{n}\BProv{P_i}{\mu_i}
 =\BProv{P}{\mu},
\]
with both disjunctions over the compatible decompositions of $\mu$.

\emph{Union.} Let $P=\opunion(P_1,\ldots,P_{n})$. By fact~(2), all rows for $\mu$ bind the same gate $g_P$, with child set $\bigcup_i\cand{P_i}{\mu}$. Thus, $\Phi_P(\mu)=\dec{g_P}=\bigvee_i\Phi_{P_i}(\mu)=\bigvee_i\BProv{P_i}{\mu}=\BProv{P}{\mu}$. If every child set is empty, $\cand{P}{\mu}=\varnothing$ and both sides are $\bot$.

\emph{Projection.} Let $P=\opproj_{Y}(P')$. For a mapping $\mu$ with $\dom(\mu)\subseteq Y$, the child set of $g_P$ is $\bigcup_{\mu':\mu'|_{Y}=\mu}\cand{P'}{\mu'}$. Hence, $\Phi_P(\mu)=\bigvee_{\mu'|_{Y}=\mu}\Phi_{P'}(\mu')=\bigvee_{\mu'|_{Y}=\mu}\BProv{P'}{\mu'}=\BProv{P}{\mu}$.

\emph{Difference.} Let $P=\opdiff(P_1,P_2)$ and let $g_P,s_P$ be the gates of Definition~\ref{def:gamma}. For every $\mu$ with $\cand{P_1}{\mu}\neq\varnothing$, $\mathit{pos}(g_P)=\cand{P_1}{\mu}$, $\mathit{ch}(s_P)=\bigcup_{\nu\sim\mu}\cand{P_2}{\nu}$ and $\mathit{neg}(g_P)=s_P$. In the subtrahend query, the group of $P_1$ binds $s_P$ before the join with $P_2$. The name $s_P$ therefore depends on the mapping of $P_1$ alone, also when $P_2$ binds a variable that the mapping of $P_1$ leaves unbound. If $\cand{P_1}{\mu}=\varnothing$, then $\Phi_P(\mu)=\bot$ and $\BProv{P}{\mu}$ implies $\BProv{P_1}{\mu}=\Phi_{P_1}(\mu)=\bot$. Otherwise $\mu$ belongs to $\mathcal I_{P_2}$ when $P_2$ is a closure atom with a variable source. If $\mu$ leaves the source unbound, the source-bound condition and Equation~\eqref{eq:world} give $\BProv{P_1}{\mu}=\bot$, and both sides are $\bot$. Otherwise every $\nu\sim\mu$ with a source takes the source $\mu(\alpha)$. The source-bound condition places $\mu(\alpha)$ in $S_{P_2}$, so Equation~\eqref{eq:atom-relative} gives $\Phi_{P_2}(\nu)=\BProv{P_2}{\nu}$ for every $\nu\sim\mu$. For any other $P_2$, the induction hypothesis gives the same equation.
Therefore,
\begin{align*}
 \Phi_P(\mu)
 &=\left(\bigvee_{g\in\cand{P_1}{\mu}}\dec g\right)
 \land\neg\left(\bigvee_{\nu\sim\mu}\bigvee_{g'\in\cand{P_2}{\nu}}\dec{g'}\right)\\*
 &=\BProv{P_1}{\mu}\land
 \neg\bigvee_{\nu\sim\mu}\BProv{P_2}{\nu}
 =\BProv{P}{\mu}.
\end{align*}
An $\opopt(P_1,P_2)$ without a filter is $\opunion(\opjoin(P_1,P_2),\opdiff(P_1,P_2))$ (Section~\ref{sec:prelim}), so the cases above cover it. A \MINUS{} follows from the reduction to $\opdiff$ below.

By fact~(2), $\cand{P}{\mu}=\{g_P\}$ for a union, projection or difference $P$, so $g_P$ denotes $\Phi_P(\mu)=\BProv{P}{\mu}$, and for $P=\opdiff(P_1,P_2)$ by the difference case, $s_P$ denotes $\bigvee_{\nu\sim\mu}\BProv{P_2}{\nu}$.
\end{proof}}

\begin{proof}[Proof of Theorem~\ref{thm:query}]
Lemma~\ref{lem:names} proves Equation~\eqref{eq:query-correct} for the root $g_Q$ of every mapping $\mu$, so the plan is sound. If $\cand{Q}{\mu}=\varnothing$, the plan produces no root for $\mu$, and Equation~\eqref{eq:candidate-formula} gives $\Phi_Q(\mu)=\bot$, which equals $\BProv{Q}{\mu}$ by the induction in the proof of Lemma~\ref{lem:names}. By Equation~\eqref{eq:world}, $\mu$ is then not an answer, so the plan is complete. The circuit is acyclic by Section~\ref{sec:correctness} and the proof of Theorem~\ref{thm:path}, because subpatterns that share a gate have the same height when filters are not counted. Every answer $\mu$ also has a root in the factored plan, because every factor is defined at the restriction of a match that extends $\mu$. Conversely, by Equation~\eqref{eq:elim-invariant}, the remaining factors are all defined at $\mu$ only when a match extends $\mu$, so Lemma~\ref{lem:factored} applies to every root of the factored plan.
\end{proof}

\paragraph{\texorpdfstring{Coverage beyond the grammar}{Coverage beyond the grammar}}
W3C $\opminus(P_1,P_2)$ removes a mapping of $P_1$ exactly when a compatible mapping of $P_2$ shares a variable with it~\cite{w3c-sparql11}. The following facts reduce $\opminus(P_1,P_2)$ to $\opdiff$ for all patterns $P_1$ and $P_2$ once their unions are moved to the top, and each fact follows from the definitions of the operators. The operator $\opminus$ distributes over a union in its first operand, and $\opminus(P_1,\opunion(P_2,P_3))=\opminus(\opminus(P_1,P_2),P_3)$. All mappings of a union-free pattern bind the same set of variables. For union-free $P_1$ and $P_2$, $\opminus(P_1,P_2)$ is $P_1$ if the mappings of $P_1$ and the mappings of $P_2$ bind disjoint sets of variables, and $\opdiff(P_1,P_2)$ otherwise. Before it moves any union, \sys applies two further facts. The pattern $\opminus(P_1,P_2)$ is $\opdiff(P_1,P_2)$ when, on every graph, every mapping of $P_1$ and every mapping of $P_2$ bind a common variable, and $P_1$ when every mapping of $P_1$ and every mapping of $P_2$ bind disjoint sets of variables. A filter $\opfilter_\varphi(P)$ mentions only variables that every mapping of $P$ binds, so $P_1$~\OPT{}~$\{P_2$~\FILTER{}~$\varphi\}$ equals $\opopt(P_1,\opfilter_\varphi(P_2))$~\cite{w3c-sparql11}. A path pattern $(\alpha,e^\diamond,c)$ with a constant target $c$ equals the closure atom $(c,(\tinv e)^\diamond,\alpha)$ when $\alpha$ is a variable, because W3C evaluates a constant target as the constant source of the inverse path~\cite{w3c-sparql11}, and $\opproj_{\varnothing}(\opfilter_{\texttt{sameTerm}(?y,c)}((\alpha,e^\diamond,?y)))$ for a fresh variable $?y$ when $\alpha$ is a constant.

\section{\texorpdfstring{Gate Names, Blank Nodes and Identifiers}{Gate Names, Blank Nodes and Identifiers}}\label{app:builder}
The hash expression $h$ of Definition~\ref{def:keys} applies \texttt{SHA256} to the encoded arguments of the name and forms the gate IRI from \texttt{urn:g:} and the first 128 bits of the digest, and $\hat\tau$ encodes the kind of a value with the digests of its lexical form, datatype and lower-cased language tag (Appendix~\ref{sup:builder}). SPARQL~1.1 only recommends lower-case digits for \texttt{SHA256}~\cite{w3c-sparql11}, so every digest passes through \texttt{LCASE}, and every engine names a gate identically. Since SPARQL~1.1 cannot sort the values of a row, the plan sorts the digests of the children of a product gate with compare-exchange steps, each a pair of \texttt{IF} expressions. The digests have fixed widths, so distinct names have distinct encodings whenever the hashed parts have distinct digests.

The proofs assume that distinct names have distinct IRIs. A leaf keeps the identifier of the triple as its IRI, and every identifier of $G_{\mathrm{reif}}$ lies outside the namespace \texttt{urn:g:} of the other gates. 
Two distinct names of a circuit with $N$ gates receive the same IRI only when the 128-bit prefixes of the digests of two distinct encodings agree or two distinct hashed strings have equal digests. Computing one name hashes fewer than $2^{64}$ strings, so in the random-oracle model~\cite{bellare1993random} the probability is below $N^2/2^{128}<3\cdot10^{-21}$ for $N=10^9$.

\paragraph{Skolem IRIs}
A key expression applies \texttt{STR} to every bound term, \texttt{STR} raises an error on a blank node, and a \BIND{} expression that raises an error leaves its variable unbound~\cite{w3c-sparql11}, so a key over a blank node drops its gate. The reified graph $G_{\mathrm{reif}}$ therefore uses fresh Skolem IRIs from a namespace that $G$ and $Q$ do not use. The filter conditions of \frag use the built-ins \texttt{STR}, \texttt{LANG}, \texttt{DATATYPE}, \texttt{isIRI}, \texttt{isLiteral}, \texttt{isBlank}, \texttt{isNumeric}, \texttt{sameTerm}, \texttt{langMatches}, \texttt{REGEX}, \texttt{IN}, \texttt{STRDT}, \texttt{CONCAT}, \texttt{CONTAINS} and \texttt{YEAR}~\cite{w3c-sparql11}. In the filter conditions of $Q$, the rewriting replaces \texttt{isBlank}, \texttt{isIRI}, \texttt{STR} and \texttt{STRDT} by expressions that return on a Skolem IRI what SPARQL returns on a blank node, and the key expressions keep the SPARQL~1.1 built-ins. The other built-ins compare terms or treat an IRI and a blank node alike. A filter therefore keeps a mapping on $G_{\mathrm{reif}}$ exactly when the filter keeps the corresponding mapping on $G$.

\paragraph{Identifier-level annotations}
Let $J$ be the identifiers in $G_{\mathrm{reif}}$, let $J_t\subseteq J$ be the nonempty set of identifiers of $t\in G$ and let the $y_i$ for $i\in J$ be independent Boolean variables with the probabilities $\Pr(y_i)$ of the identifiers. The substitution $\sigma$ maps $x_t$ to $\bigvee_{i\in J_t}y_i$. For an identifier world $\omega\subseteq J$, let $W_\omega=\{t\in G\mid J_t\cap\omega\ne\varnothing\}$, and let $\mathsf{val}_\omega$ make $y_i$ true exactly when $i\in\omega$. Since substitution preserves conjunction, disjunction and negation, every Boolean function $f$ over $X$ satisfies
\begin{equation}
 \mathsf{val}_\omega(\sigma(f))=\mathsf{val}_{W_\omega}(f).
 \label{eq:identifier-substitution}
\end{equation}
The rows of a matching triple $t$ bind the leaves $y_i$ for $i\in J_t$, whose disjunction is $\sigma(x_t)$, and the inductions in the proofs of Theorem~\ref{thm:path} and Lemma~\ref{lem:names} carry over under $\sigma$, so the root of $\mu$ denotes $\sigma(\BProv{Q}{\mu})$. By Equations~\eqref{eq:identifier-substitution} and~\eqref{eq:world}, the root is true at $\omega$ exactly when $\mu\in\supp(\sem{Q}{W_\omega})$, so its weighted model count is the marginal probability of $\mu$ over the worlds $W_\omega$.
\section{Tests and the Reimplementation of NPCS}\label{app:artifacts}
Tests compare the probabilities of \sys with an enumeration of the possible worlds on query shapes that cover~\frag. For the comparison with NPCS, we reimplemented the rewriting rules of NPCS. The published NPCS rewriter is a compiled JAR that returns no provenance or provenance without triples for queries with \OPT{}, \UNION{}, \kw{distinct} or \FILTER{} and evaluates \MINUS{} as \OPT{}. On the 138 basic graph pattern queries of the WatDiv workload of NPCS, our reimplementation returns the same rewritten queries as the JAR up to variable names and operand order.

\section{Proofs in Detail}\label{sup:proofs}

\subsection{Factored basic graph patterns}\label{sup:factored}
Appendix~\ref{app:factored} proves Lemma~\ref{lem:factored} for the base factors of Definition~\ref{def:factored}. When a triple pattern of $Q$ has a constant subject or object, the factored plan computes all base factors with one query that matches the whole basic graph pattern. Every base factor is therefore defined only at restrictions of these matches.

The restriction of the base factors drops a mapping only when no match of the basic graph pattern extends the mapping. Every total binding ${\rho}$ that extends a dropped mapping then has ${\rho}({\tp_l})\notin G$ for some $l$, and therefore $F_l(\rho|_{\var(\tp_l)})=\bot$ and the conjunction in Equation~\eqref{eq:flat-formula} is $\bot$ at $\rho$. The dropped mappings contribute only $\bot$ to Equation~\eqref{eq:flat-formula}, so the restriction leaves the equation unchanged.

For the bound of Section~\ref{sec:factored}, the induced width gives $|U_z|\le w$ for every eliminated $z$. Each factor $H_z$ is defined at most at $|\adom(G)|^w$ mappings, while its product gates range over at most $|\adom(G)|^{w+1}$ assignments to $U_z\cup\{z\}$. Since $|\adom(G)|\le3|G|$ and there are $O(m)$ base factors and elimination steps, the elimination steps create $O(m(3|G|)^{w+1})$ gates, which is $O(m|G|^{w+1})$ for fixed $w$. The roots add one product gate and one sum gate per answer.

\subsection{Gate names in the proof of Lemma~\ref{lem:names}}\label{sup:names}
The proof of Lemma~\ref{lem:names} in Appendix~\ref{app:proofs} uses three facts about the names, which we prove here: (1)~two product gates have the same IRI if and only if their lists of children are equal up to order, (2)~two sum gates, or two difference gates, have the same IRI if and only if they have the same letter, the same mapping and subpatterns equal up to the identities of Definition~\ref{def:ids}, and (3)~the IRI of a gate depends only on the values that the row binds. Under the assumption, two gates have the same IRI exactly when they have the same name. (1)~The name of a product gate is $(\texttt{T},\operatorname{sort}(c_1,\ldots,c_n))$ for the IRIs $c_1,\ldots,c_n$ of its children. Two product gates with the same IRI have the same name by the assumption, so their lists of children are equal up to order. Conversely, two such lists have the same hashes in ascending order, and the IRI depends only on these hashes. (2)~The name of a sum gate is $(L,\theta(P),\tau(a_1),\ldots,\tau(a_k))$ for its letter $L$, its subpattern $P$ and the values $a_j$ of the variables $?x_1<\cdots<?x_k$ in scope at $P$ under its mapping $\mu$, with $a_j=\mathsf{unbound}$ when $\mu$ leaves $?x_j$ unbound. Since $\theta(P)$ determines $P$ up to the identities of Definition~\ref{def:ids}, which preserve Boolean provenance, and $\tau$ is injective, two names are equal exactly when the letters, the mappings and the subpatterns up to these identities are equal. Difference gates have the letter \texttt{M} and the same argument. (3)~A key expression applies only the functions of Section~\ref{sup:builder} to the values of a row, SPARQL~1.1 fixes their values on IRIs and literals, and $G_{\mathrm{reif}}$ contains no blank nodes. The IRI of a gate therefore depends only on the values that the row binds.

\section{\texorpdfstring{Gate Names, Collisions and Identifiers in Detail}{Gate Names, Collisions and Identifiers in Detail}}\label{sup:builder}
The hash expression $h$ of Definition~\ref{def:keys} is
\[
\begin{array}{@{}l@{}}
h(\texttt{T},d_1,\ldots,d_n)=\texttt{CONCAT("urn:g:t:",}\\
\qquad\delta_{128}(\texttt{CONCAT("T|",}\,d_1,\texttt{"|"},\ldots,\texttt{"|"},d_n\texttt{)})\texttt{)},\\[2pt]
h(L,\theta,a_1,\ldots,a_k)=\texttt{CONCAT("urn:g:",}\,\texttt{LCASE(}L\texttt{)}\texttt{,":",}\\
\qquad\delta_{128}(\texttt{CONCAT(}L\texttt{,"@",}\,\delta(\theta),\eta_1,\ldots,\eta_k\texttt{)})\texttt{)},
\end{array}
\]
where $L$ is one of the letters \texttt{U}, \texttt{P}, \texttt{A}, \texttt{S} and \texttt{M}, $\delta(\cdot)$ is \texttt{LCASE(SHA256(}$\cdot$\texttt{))} and $\delta_{128}(\cdot)$ is \texttt{SUBSTR(}$\delta(\cdot)$\texttt{, 1, 32)} for a string expression, and $\eta_j$ is $\delta(\texttt{CONCAT("}x_j\texttt{=",}\,a_j\texttt{)})$ for the name $x_j$ of $?x_j$. The children enter $h(\texttt{T},\ldots)$ as their digests $\delta(\texttt{STR(?}c_i\texttt{)})$ in ascending order, and $\hat\tau(?x)$ encodes $\tau$ of the value of $?x$:
\[
\begin{array}{@{}l@{}}
\texttt{IF(!BOUND(?x), \textquotedbl u\textquotedbl,}\\
\quad\texttt{IF(isIRI(?x), CONCAT(\textquotedbl i\textquotedbl,}\,\delta(\texttt{STR(?x)})\texttt{),}\\
\quad\texttt{IF(isBlank(?x), CONCAT(\textquotedbl b\textquotedbl,}\,\delta(\texttt{STR(?x)})\texttt{),}\\
\quad\texttt{IF(isLiteral(?x), CONCAT(\textquotedbl l\textquotedbl,}\,\delta(\texttt{STR(?x)}),\\
\qquad \delta(\texttt{STR(DATATYPE(?x))}),\,\delta(\texttt{LCASE(LANG(?x))})\texttt{),}\\
\quad\texttt{CONCAT(\textquotedbl t\textquotedbl,}\,\delta(\texttt{STR(?x)})\texttt{)))))}.
\end{array}
\]
The letter \texttt{t} marks a quoted triple of SPARQL-star~\cite{hartig2021rdfstar}. On $G_{\mathrm{reif}}$, every value is an IRI, a literal or a quoted triple, so the branch for blank nodes is never taken. SPARQL~1.1 only recommends lower-case digits for \texttt{SHA256}~\cite{w3c-sparql11}, and $\delta$ therefore applies \texttt{LCASE}, so every engine names a gate identically. SPARQL~1.1 offers no function that sorts the values of a row, and the plan sorts the digests of the children of a product gate with compare-exchange steps of two \texttt{IF} expressions each. The digests have fixed widths, so names whose hashed parts have distinct digests have distinct encodings.

The proofs assume that distinct names have distinct IRIs. A leaf keeps the identifier of the triple as its IRI, and every identifier of $G_{\mathrm{reif}}$ lies outside the namespace \texttt{urn:g:} of the other gates. For a circuit with $N$ gates, let $M$ be the number of distinct strings that the plan hashes to compute the names. Two distinct names receive the same IRI only when the digests of two distinct encodings agree in the first 128 bits or two of the $M$ strings have equal digests. In the random-oracle model~\cite{bellare1993random}, the probability of either event is at most $N(N-1)/2^{129}+M(M-1)/2^{257}$. Computing one name hashes fewer than $2^{64}$ strings, so $M<2^{64}N$ and the probability is below $N^2/2^{128}$. For $N=10^9$ gates, $N^2/2^{128}<3\cdot10^{-21}$.

\paragraph{Identifier-level annotations}
Let $J$ be the identifiers in $G_{\mathrm{reif}}$, let $J_t\subseteq J$ be the nonempty set of identifiers of $t\in G$, and let $y_i$ be a Boolean variable with the probability $\Pr(y_i)$ of the identifier $i$ for every $i\in J$. The variables $y_i$ are independent. In Section~\ref{sec:prelim}, the sets $J_t$ are disjoint singletons. Multiple Wikidata statements give several identifiers for one triple, whereas the RDF triples of a TPC-H row share its identifier. The substitution $\sigma$ maps $x_t$ to $\bigvee_{i\in J_t}y_i$. For an identifier world $\omega\subseteq J$, let $W_\omega=\{t\in G\mid J_t\cap\omega\ne\varnothing\}$, and let the valuation $\mathsf{val}_\omega$ make $y_i$ true exactly when $i\in\omega$. Since substitution preserves conjunction, disjunction and negation, every Boolean function $f$ over $X$ satisfies
\begin{equation}
 \mathsf{val}_\omega(\sigma(f))=\mathsf{val}_{W_\omega}(f).
 \label{sup:identifier-substitution}
\end{equation}
For a matching triple $t$, the encoded rows bind the leaves $\{y_i\mid i\in J_t\}$, whose disjunction is $\sigma(x_t)$, and the inductions in the proofs of Theorem~\ref{thm:path} and Lemma~\ref{lem:names} carry over to the functions under $\sigma$. The root of $\mu$ therefore denotes $\sigma(\BProv{Q}{\mu})$. By Equations~\eqref{sup:identifier-substitution} and~\eqref{eq:world}, the root is true at $\omega$ exactly when $\mu\in\supp(\sem{Q}{W_\omega})$. The weighted model count of the root is the sum of $\prod_{i\in\omega}\Pr(y_i)\prod_{i\in J\setminus\omega}(1-\Pr(y_i))$ over the sets $\omega\subseteq J$ at which the root is true, that is, the marginal probability of $\mu$ over the worlds $W_\omega$. Shared identifiers correlate the RDF triples of a row, but the $y_i$ stay independent.

\section{\texorpdfstring{The Factored Plan on a Hub}{The Factored Plan on a Hub}}\label{app:full-hub}
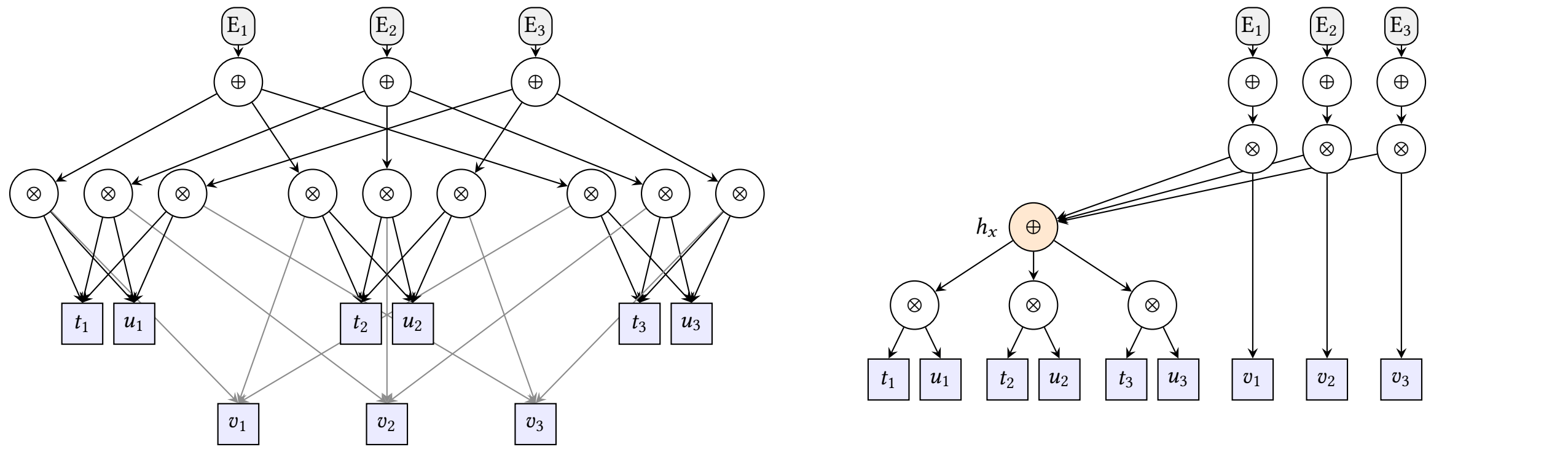
\begin{figure*}[htbp]
\centering
\pgfdeclarelayer{edges}\pgfsetlayers{edges,main}
\begin{tikzpicture}[>=stealth,
  ans/.style={draw, rounded corners, fill=gray!12, inner sep=2pt, minimum height=5mm, font=\normalsize\bfseries},
  gate/.style={draw, circle, fill=white, minimum size=6.5mm, inner sep=0pt, font=\normalsize},
  leaf/.style={draw, fill=blue!8, minimum size=5.5mm, inner sep=1pt, font=\normalsize\bfseries},
  shgate/.style={draw, circle, fill=orange!18, minimum size=6.5mm, inner sep=0pt, font=\normalsize},
  scale=0.8, transform shape]
  \foreach \j/\x in {1/3.25, 2/5.25, 3/7.25} {
    \node[ans] (za\j) at (\x,4.75) {$\mathrm{E}_\j$};
    \node[gate] (ra\j) at (\x,4.0) {$\oplus$};
  }
  \foreach \i/\x in {1/1.5, 2/5.25, 3/9.0} {
    \node[leaf] (ta\i) at ({\x-0.35},0.75) {$t_\i$};
    \node[leaf] (ua\i) at ({\x+0.35},0.75) {$u_\i$};
    \foreach \j in {1,2,3} \node[gate] (pa\i\j) at ({\x+1.0*(\j-2)},2.5) {$\otimes$};
  }
  \foreach \j/\x in {1/3.25, 2/5.25, 3/7.25} \node[leaf] (va\j) at (\x,-0.6) {$v_\j$};
  \node[font=\normalsize] at (5.25,-1.45) {(a) flat plan: 12 gates, 36 edges};
  \begin{scope}[xshift=12.0cm]
  \foreach \i in {1,2,3} {
    \node[leaf] (tb\i) at ({1.6*(\i-1)},0) {$t_\i$};
    \node[leaf] (ub\i) at ({0.7+1.6*(\i-1)},0) {$u_\i$};
    \node[gate] (qb\i) at ({0.35+1.6*(\i-1)},1.0) {$\otimes$};
  }
  \node[shgate] (hx) at (1.95,2.05) {$\oplus$};
  \node[font=\normalsize, anchor=east] at (1.6,2.05) {$h_x$};
  \foreach \j/\x in {1/4.9, 2/5.9, 3/6.9} {
    \node[leaf] (vb\j) at (\x,0) {$v_\j$};
    \node[gate] (sb\j) at (\x,3.1) {$\otimes$};
    \node[gate] (rb\j) at (\x,4.0) {$\oplus$};
    \node[ans] (zb\j) at (\x,4.75) {$\mathrm{E}_\j$};
  }
  \node[font=\normalsize] at (3.45,-1.45) {(b) factored plan: 10 gates, 18 edges};
  \end{scope}
  \begin{pgfonlayer}{edges}
  \foreach \j in {1,2,3} {
    \draw[->] (za\j) -- (ra\j);
    \foreach \i in {1,2,3} {
      \draw[->] (ra\j) -- (pa\i\j);
      \draw[->] (pa\i\j) -- (ta\i.north);
      \draw[->] (pa\i\j) -- (ua\i.north);
      \draw[->, black!45] (pa\i\j) -- (va\j.north);
    }
    \draw[->] (zb\j) -- (rb\j); \draw[->] (rb\j) -- (sb\j);
    \draw[->] (sb\j) -- (hx); \draw[->] (sb\j) -- (vb\j.north);
  }
  \foreach \i in {1,2,3} {
    \draw[->] (hx) -- (qb\i);
    \draw[->] (qb\i) -- (tb\i.north); \draw[->] (qb\i) -- (ub\i.north);
  }
  \end{pgfonlayer}
\end{tikzpicture}
\caption{Circuits of the running query over $G_{\mathrm{hub}}$. The gray edges in (a) lead to the leaves $v_j$, and the three answers in (b) share the orange sum gate $h_x$.}
\Description{Two circuits for the answers E1, E2 and E3. In (a), every root has three product gates, one for each pair t_i and u_i together with v_j, so nine product gates reach the shared leaves. The product gates are grouped by the pair t_i and u_i, and the edges to v_j are gray. In (b), one sum gate h_x over the three products of t_i and u_i is shared, and every root has one product gate over h_x and v_j.}
\label{fig:hub}
\end{figure*}

For $i,j\in\{1,2,3\}$, let $G_{\mathrm{hub}}$ contain the triples
\[
\begin{array}{@{}r@{}l@{}}
 t_i&{}=(\mathrm{Aspirin},\mathrm{interactsWith},\mathrm{D}_i),\\
 u_i&{}=(\mathrm{D}_i,\mathrm{interactsWith},\mathrm{Metformin}),\\
 v_j&{}=(\mathrm{Metformin},\mathrm{interactsWith},\mathrm{E}_j).
\end{array}
\]
The names $t_i$, $u_i$ and $v_j$ also denote the leaves of their triples. Over $G_{\mathrm{hub}}$, the running query returns $\mathrm{E}_1$, $\mathrm{E}_2$ and $\mathrm{E}_3$, and every answer $\mathrm{E}_j$ has the three derivations $(t_i,u_i,v_j)$. The flat plan of Definition~\ref{def:gamma} creates one product gate per derivation, so the circuit in Figure~\ref{fig:hub}(a) has nine product gates with three edges each below three roots. The factored plan of Definition~\ref{def:factored} eliminates $?x$ first, and $H_x$ maps Metformin to $h_x=\oplus(\otimes(t_1,u_1),\otimes(t_2,u_2),\otimes(t_3,u_3))$. Eliminating $?y$ then yields the factor $H_y$ that maps $\mathrm{E}_j$ to $\oplus(\otimes(h_x,v_j))$, so the three answers share $h_x$ in Figure~\ref{fig:hub}(b).

For $k$ drugs between Aspirin and Metformin and $l$ drugs after Metformin, the flat circuit has $kl+l$ gates other than leaves and $4kl$ edges. After the removal of non-root sum gates with one child, the factored circuit has $k+2l+1$ gates and $3(k+l)$ edges. For $k=l=3$, the flat circuit has 12 gates and 36 edges, and the factored circuit 10 gates and 18 edges. \sys emits these circuits on $G_{\mathrm{hub}}$.

\section{\texorpdfstring{Transitive Closures in Detail}{Transitive Closures in Detail}}\label{app:full-paths}
Let $D_G$ contain the subjects and objects occurring in $G$. For every closure-free expression $e$ and $u,v\in D_G$, we construct a gate $b^e_{uv}$ by recursion on $e$. A construction with no child uses $g_\bot$. For an IRI $p$,
\[b^p_{uv}=\oplus\bigl(\{x_t\mid t=(u,p,v)\in G\}\bigr).\]
Inverse and alternative use
\[
 b^{\tinv e}_{uv}=b^e_{vu}\,\,\text{and}\,\,
 b^{e_1\talt e_2}_{uv}=\oplus\bigl(\{b^{e_1}_{uv},b^{e_2}_{uv}\}\setminus\{g_\bot\}\bigr).
\]
For sequence, let
\begin{equation}
 b^{e_1\tseq e_2}_{uv}=\oplus\!\left(
 \left\{
  \otimes\bigl(\{b^{e_1}_{uw},b^{e_2}_{wv}\}\bigr)
  \middle|
  \begin{array}{l}
   w\in D_G,\ b^{e_1}_{uw}\ne g_\bot,\\[-2pt]
   b^{e_2}_{wv}\ne g_\bot
  \end{array}
 \right\}
 \right).
 \label{eq:closurefree-gates}
\end{equation}
Gate keys contain the syntax node of $e$ and the node pair. Inverse reuses the gate $b^e_{vu}$, so the gate keys mention only IRI, sequence and alternative nodes. These equations define the global closure-free circuit $C_e$. The equations index a gate by every node pair, so Lemma~\ref{lem:closurefree} counts the gates of $C_e$ over the whole graph. A plan materializes only the gates other than $g_\bot$ that a root reaches.

\begin{lemma}\label{lem:closurefree}
For every closure-free $e$, $u,v\in D_G$ and world $W\subseteq G$,
\[
 \val{W}(\dec{b^e_{uv}})=\top
 \,\,\Longleftrightarrow\,\,
 (u,v)\in\relpath{e}{W}.
\]
The circuit $C_e$ is acyclic and has $O(|e|\,|D_G|^3)$ gates.
\end{lemma}

\begin{proof}[Proof of Lemma~\ref{lem:closurefree}]
We use structural induction on $e$. The base case is an IRI $p$, and the induction hypothesis is the claim for the direct subexpressions of $e$ at all $u,v\in D_G$ and all $W\subseteq G$.

\emph{IRI.} Let $e=p$. The semantics of Section~\ref{sec:prelim} gives
\begin{align*}
 \val{W}(\dec{b^p_{uv}})=\top
 &\Longleftrightarrow
 \bigvee_{t=(u,p,v)\in G}\val{W}(x_t)=\top\\
 &\Longleftrightarrow (u,p,v)\in W\\
 &\Longleftrightarrow (u,v)\in\relpath{p}{W}.
\end{align*}
The empty child set yields $\bot$ on both sides.

\emph{Inverse.} Let $e=\tinv e_1$. By construction and induction,
\begin{align*}
 \val{W}(\dec{b^{\tinv e_1}_{uv}})=\top
 &\Longleftrightarrow \val{W}(\dec{b^{e_1}_{vu}})=\top\\
 &\Longleftrightarrow (v,u)\in\relpath{e_1}{W}\\
 &\Longleftrightarrow (u,v)\in\relpath{\tinv e_1}{W}.
\end{align*}

\emph{Alternative.} Let $e=e_1\talt e_2$. Since $\dec{g_\bot}=\bot$, omitting $g_\bot$ does not change the disjunction. The two induction hypotheses yield
\begin{align*}
 \val{W}(\dec{b^{e_1\talt e_2}_{uv}})=\top
 &\Longleftrightarrow
 \val{W}(\dec{b^{e_1}_{uv}})=\top\\[-1pt]
 &\hspace{24pt}\lor \val{W}(\dec{b^{e_2}_{uv}})=\top\\
 &\Longleftrightarrow
 (u,v)\in\relpath{e_1}{W}\lor(u,v)\in\relpath{e_2}{W}\\
 &\Longleftrightarrow (u,v)\in\relpath{e_1\talt e_2}{W}.
\end{align*}

\emph{Sequence.} Let $e=e_1\tseq e_2$. Equation~\eqref{eq:closurefree-gates}, the semantics of Section~\ref{sec:prelim} and the induction hypotheses give
\begin{align*}
 \val{W}(\dec{b^{e_1\tseq e_2}_{uv}})=\top
 &\Longleftrightarrow \exists w\in D_G:\
 \val{W}(\dec{b^{e_1}_{uw}})=\top\\[-1pt]
 &\hspace{52pt}\land\val{W}(\dec{b^{e_2}_{wv}})=\top\\
 &\Longleftrightarrow \exists w\in D_G:\
 (u,w)\in\relpath{e_1}{W}\\[-1pt]
 &\hspace{52pt}\land(w,v)\in\relpath{e_2}{W}\\
 &\Longleftrightarrow (u,v)\in\relpath{e_1\tseq e_2}{W}.
\end{align*}
If no midpoint satisfies the first line, the outer disjunction is empty and both sides are false. The cases above exhaust the grammar.

Let $d=|D_G|$. Each IRI or alternative syntax node creates $O(d^2)$ gates. Each sequence node creates at most one sum gate per node pair and one product gate per node triple $(u,w,v)$, so the count per sequence node is $O(d^3)$. The syntax tree has $O(|e|)$ nodes, so $C_e$ has $O(|e|d^3)$ gates. Every edge leads to a leaf, to a gate of a proper subexpression or from the sum gate of a sequence to one of its product gates. The children of a product gate of a sequence are gates of the two subexpressions. Ordering the gates by expression depth and placing the product gates of a sequence below its sum gates therefore gives a topological order.
\end{proof}

Let $E_e$ contain the pairs $(u,v)\in D_G^2$ with $b^e_{uv}\ne g_\bot$. The set $V_s$ of Section~\ref{sec:paths} contains $s$ and the nodes that $E_e$ reaches from $s$, and $E_s=E_e\cap(V_s\times V_s)$. Only part of $C_e$ is ever needed. The levels of Section~\ref{sec:paths} use $b^e_{uv}$ for $(u,v)\in E_s$ alone, and expanding one such gate involves $b^{e'}_{uw}$ only for a subexpression $e'$ of $e$ and a midpoint $w$ that a prefix of $e$ reaches from $u$. \sys therefore emits these gates only on the matches of $e$ that start in $V_s$. Choosing one part at every alternative node turns $e$ into finitely many sequences, and for each sequence one \CONSTRUCT{} query matches it from the nodes of $V_s$ and emits the gates of the syntax nodes that the sequence passes through. For $e=p$, these are the $O(|E_s|)$ gates $b^p_{uv}$ with $(u,v)\in E_s$. Lemma~\ref{lem:closurefree} and the count of level gates in the proof of Theorem~\ref{thm:path} give the global total $O(|e|\,|D_G|^3+n(n+|E_s|))$. For $e=p$, the total is $O(n(n+|E_s|))$.

\sys first collects $V_s$ with \SELECT{} queries that each follow one $e$-step from the nodes found so far, and then emits the gates $b^e_{uv}$ with $u\in V_s$ together with level~$1$. \sys keeps one row per gate of level~$i$ with the source and the target of the gate, and issues \CONSTRUCT{} queries for the product and sum gates of level~$i+1$. Every level gate records the level with \texttt{c:rlvl}, the digest of $e^\diamond$ with \texttt{c:rpath}, the source with \texttt{c:rfrom} and the target with \texttt{c:rto}. \sys derives the rows from the level, the source and the target. Gate keys contain $e^\diamond$, $s$, $i$ and the target, so every edge points to a lower level and the two closures of one expression build separate circuits. The key of the closure gate $R(v)$ contains $e^\diamond$, $s$ and $v$ alone. The base gates $b^e_{uv}$ and the zero-length gate carry distinguished levels of their own, so no round composes the zero-length gate. A last \CONSTRUCT{} query emits the closure gates $R(v)$ together with the roots or the rows of the closure atom, which take the zero-length gate for the pair $(s,s)$ when $\diamond=*$. For a closure that forms the whole query, the level queries receive the rows of level~$i$ in a \VALUES{} clause and only read the endpoint. When the rows exceed the size of one request, \sys splits them over several queries and gives each query the base rows that start at the targets of its rows. The merged results equal those of one query.

For a closure atom inside a larger pattern, \sys inserts the rows into the private namespace under two predicates, and the level queries read the rows from the endpoint. The names of the two predicates contain the run, the digest of $e^\diamond$, the source and the level, and therefore keep the rows of different runs, closure atoms, sources and levels apart. The digest also separates $e^+$ from $e^*$, so only the star construction reads the zero-length row that it inserts. A source variable from an enclosing pattern invokes the same construction for each source binding, and the query that emits $o_s$ matches the triples whose subject or object is the source. RDF4J parses a closure atom whose source is its target variable, $(?x,e^\diamond,?x)$, as the atom $(?x,e^\diamond,?t)$ with a fresh variable $?t$ under the filter $\texttt{sameTerm}(?x,?t)$. \sys passes the source binding through this filter and keeps the column of $?t$ in the relation, so the filter keeps the pairs $(s,s)$.

The Bellman--Ford circuit of Fan et al.~\cite{fan2025circuits} carries $r_i(v)$ into level $i+1$, so one level accumulates every length. In the endpoint, the carried gates republish a pair at every later level as a sum gate with one input. The level is also part of a key, so the products $\otimes(\{r_i(u),b^e_{uv}\})$ and $\otimes(\{r_{i+1}(u),b^e_{uv}\})$ have different children and are distinct gates. Both products cover the walks of length at most $i+1$ that end with the step $(u,v)$. Exact-length levels hold each walk at one level and avoid both costs. The construction stops at the first empty level or after $n$ levels (Section~\ref{sec:paths}). The bound $n$ and the emptiness of a level depend only on $Q$ and $G$, so \sys stops at the same $\lambda$ and emits the same gates on every engine.
\section{\texorpdfstring{Implementation Details}{Implementation Details}}\label{app:full-impl}
\paragraph{Gate names}
\sys computes $\delta(\theta(P))$ once for every subpattern $P$, and the tag of the outer projection serializes the parsed query. The query fixes the number $n$ of children of a product gate, and the plan sorts their digests with the $n(n-1)/2$ compare-exchange steps of a bubble-sort network. Each step is the pair \texttt{IF(?a <= ?b, ?a, ?b)} and \texttt{IF(?a <= ?b, ?b, ?a)}.

\paragraph{\texorpdfstring{Roots}{Roots}}
The prefix \texttt{c:} stands for \texttt{urn:circuit:}. The query that emits a root $r$ also emits the triple $r$~\texttt{c:answerRoot}~$\mathit{vs}$ with a literal $\mathit{vs}$ that lists the projected variables. For every projected variable $?y$ that the mapping of $r$ binds to a value $v$, the query also emits the triple $r$~\texttt{c:bind:}$\mathit{hx}$~$v$. The suffix $\mathit{hx}$ is the hexadecimal encoding of the UTF-8 bytes of the name of $?y$, and $\mathit{vs}$ consists of \texttt{vars:} followed by the comma-separated suffixes of all projected variables. The plan~\eqref{eq:plan} writes \texttt{z} in place of its encoding \texttt{7a}. \sys reads the mapping of every root from these triples.

\paragraph{Skolem IRIs}
A key over a blank node drops its gate (Appendix~\ref{app:builder}) because a template omits every triple with an unbound variable. The label of a blank node is scoped to one query result~\cite{w3c-sparql11}, so two queries can return different labels for one blank node. In the filter conditions of $Q$, the expressions that replace \texttt{isBlank}, \texttt{isIRI}, \texttt{STR} and \texttt{STRDT} test whether an IRI is a Skolem IRI. On a Skolem IRI, \texttt{isBlank} then returns true and \texttt{isIRI} false, and \texttt{STR} and \texttt{STRDT} raise the error that SPARQL raises for a blank node. Every Skolem IRI in an answer is mapped back to its blank node.

\paragraph{The implemented plan}
\sys builds the queries of Definition~\ref{def:gamma} with two changes. The query of the operator above a join emits the product gates of the join, which has no query of its own. The query contains the \WHERE{} clause of the join, and its template emits the product gate of every row of the join with the edges to the children of the gate. Every product gate that a root reaches is therefore in the circuit. The pattern $\rwp{P}{g_P}$ of a union $P$ ends with the \BIND{} of $g_P$. On RDF4J 4.2.1 and 5.3.2~\cite{rdf4j}, we observed that a join of two groups is evaluated as a hash join on the shared variables when one group contains a \BIND{}. The hash join drops every row that leaves a shared variable unbound. SPARQL makes the row compatible with every row that agrees on the bound variables of the row. Over the triples \texttt{:s~:p0~:a}, \texttt{:s~:p0~:b}, \texttt{:s~:p1~:c} and \texttt{:s~:p2~:d}, the pattern
{\begin{verbatim}
  { {?x :p0 ?v2}
    { {?x :p1 ?v1 . ?x :p2 ?v2} UNION {?x :p1 ?v1}
      BIND("R" AS ?g) } }
\end{verbatim}}
\noindent returns 0 rows on both versions, and the correct result has 2 rows.{} Before applying Definition~\ref{def:gamma}, \sys therefore rewrites a join or a difference whose operand has a union with a branch that leaves unbound a variable of the other operand. Join and the left operand of a difference distribute over union, and $\opdiff(P_1,\opunion(P_2,P_3))=\opdiff(\opdiff(P_1,P_2),P_3)$. The rewritten query therefore has the answers and the Boolean provenance of the original query. For the RDF-star annotation, standard reification and named graphs, \sys matches a triple pattern together with its encoding. The matches remain those of Definition~\ref{def:gamma}, because $G_{\mathrm{reif}}$ contains the triples of $G$ with Skolem IRIs in place of blank nodes. \sys also adds \kw{distinct} to the subquery of a projection. The keyword removes duplicate solutions and leaves the rows and the circuit unchanged.

\paragraph{Factored basic graph patterns}
Every row of a factor is an RDF node in the private namespace that stores the mapping and the gate of the row. The sum gates of a factor carry the prefix \texttt{urn:g:s:} and hash a tag together with the bindings of the scope. The tag of a base factor combines a digest of the basic graph pattern with the index of its triple pattern, and the tag of any other factor combines the digest with the eliminated variable and the step. The subtrahend sum gates of Definition~\ref{def:keys} carry the same prefix, and only their hashed strings begin with \texttt{S@}. A base factor maps a row to the sum gate over the identifiers of its triple. One \INSERT{} makes the rows of a factor available to the next query. At every step, \sys eliminates a variable $z$ with the smallest $|U_z|$ and breaks ties by the name of the variable. The factored plan multiplies the factors of $\mathcal A_z$ two at a time. Pairwise multiplication changes the nesting of the product gates and preserves their Boolean functions and the bound of Section~\ref{sec:factored}. \sys uses the flat plan in three cases. In the first case, the basic graph pattern has at most one triple pattern, so every derivation consists of at most one triple. The other two cases concern a basic graph pattern whose subjects and objects are variables, and every base factor then contains every match of its triple pattern in $G$. In the second case, the query projects every variable, so every answer has one derivation. In the third case, a triple pattern has a variable predicate, and its base factor contains every triple of $G$.
\paragraph{Compilation}
Before compilation, \sys replaces every sum gate that has one child and is not a root by its child. For d4, \sys compiles the Tseitin encoding~\cite{tseitin1968complexity} of the circuit below the root, a CNF with one auxiliary variable per gate. Weight~$1$ on both literals of each auxiliary variable keeps the WMC~\cite{fierens2015problog2,sang2005wmc}.
\section{\texorpdfstring{Builder Independence}{Builder Independence}}\label{app:full-builder}
Definition~\ref{def:ids} derives every gate name from the operator, the
pattern tag and either the bindings of the row or the IRIs of the children, all of which $Q$ and $G_{\mathrm{reif}}$ determine. Every engine that follows the SPARQL~1.1 semantics and evaluates the plan over $G_{\mathrm{reif}}$ therefore returns the same circuit.

For a subpattern $P$, let $\rewrwit(P)$ be the \SELECT{} queries obtained from
each \CONSTRUCT{} query of Definition~\ref{def:gamma} by replacing its template
with the projection of $\bar v_P$ and the child gate variables of that case. A union
therefore has one such query per branch and a difference two, and $\bar v_P$ is $\bar v_{P_1}$ for a difference. The witness query of $q^-_P$ projects $s_P$, which the group of $P_1$ binds before the join with $P_2$, in place of $\bar v_{P_1}$, because a variable unbound in the row of $P_1$ can take a value of $P_2$. We
call their results the \emph{witness relations} of $P$.

Applying the templates of Definition~\ref{def:gamma} to the witness relations of every subpattern of $Q$ yields the circuit of the plan $\rewr(Q)$ with the same gate names, kinds and child sets. The plan and the witness relations give the key expressions the same arguments: the pattern tag and the binding tuple for $\oplus$- and $\ominus$-gates, and the child gates for $\otimes$-gates. A witness relation returns the binding tuple verbatim, the gate $s_P$ of a subtrahend query, and the child gates as its gate columns. In both cases, the children of one name are those of the rows that bind the name.

Emitting the gates inside the engine and building them from the witness relations differ in cost. Let $R$ be the number of rows that a
plan enumerates, $N$ the number of distinct gates and $E$ the number of edges of its circuit. Emission inside the engine materializes $\Theta(N+E)$ triples, because set semantics merges a gate the
moment it is emitted a second time. A witness relation instead returns $\Theta(mR)$ terms for $m$ patterns. The two coincide only when $N+E$ is $\Theta(mR)$. The flat
plan of a basic graph pattern gives every row a product gate with at most $m$ children and one edge from the root, so $E\le(m+1)R$, and
$N+E$ is $\Theta(mR)$ when distinct rows match distinct sets of triples. The factored plan makes $E$ small even when the flat plan enumerates many rows, and
the gain holds only while its intermediate factors stay in the endpoint that evaluates the joins.
\section{\texorpdfstring{Additional Experiments}{Additional Experiments}}\label{app:experiments}
This section gives the results of Section~\ref{sec:eval} for all engines and the results that Section~\ref{sec:eval} leaves out. On WatDiv, the endpoint and the remaining steps each have 600\,s with 10.9 million triples and 3000\,s with 109 million triples. On gMark, the endpoint has 5000\,s. On Wikidata, a run has 600\,s from the rewriting of the query to the probabilities. On TPC-H, a run has 3000\,s from the rewriting of the query to the probabilities.

\subsection{WatDiv}\label{app:exp:watdiv}
\begin{table}[htbp]
\caption{End-to-end time (s) on WatDiv (RDF-star).}
\label{tab:watdiv-full}
\small
\begin{tabular*}{\columnwidth}{@{}l@{\extracolsep{\fill}}rrrrrr@{}}
\toprule
 & \multicolumn{2}{c}{GraphDB} & \multicolumn{2}{c}{Fuseki} & \multicolumn{2}{c}{Oxigraph}\\
\cmidrule(lr){2-3}\cmidrule(lr){4-5}\cmidrule(l){6-7}
Triples & 10.9M & 109M & 10.9M & 109M & 10.9M & 109M\\
\midrule
NPCS & 0.46 & 0.80 & 0.49 & 0.87 & 0.69 & 11.89\\
Flat plan & 0.87 & 1.11 & 0.82 & 1.43 & 1.09 & 12.88\\
Factored plan & 1.34 & 2.20 & 1.15 & 3.29 & 1.15 & 1.53\\
\bottomrule
\end{tabular*}
\end{table}
NPCS reports an overhead of 20 to 30\% over the original queries on GraphDB with a cold cache~\cite{asma2024npcs}. Our runs follow a warm-up run, the plain query runs on $G$ and NPCS on $G_{\mathrm{reif}}$, and with 10.9 million triples on GraphDB, NPCS takes 1.9 times as long as the reified query, which takes 8.7 times as long as the plain query (Section~\ref{sec:eval:overhead}).

Table~\ref{tab:watdiv-full} gives the end-to-end time over the queries that NPCS and both plans complete. At the endpoint, the flat plan costs 2.5 to 2.6 times NPCS on GraphDB and the factored plan 4.3 to 5.1 times. Fuseki and Oxigraph follow GraphDB, except on Oxigraph with 109 million triples. There, the reified query already takes 9.1 times as long as the plain query. The factored plan builds every factor in a separate query and takes 0.2 times as long as NPCS at the endpoint on the 19 templates of basic graph patterns that it completes. End to end, the factored plan returns all probabilities in 1.53\,s against 11.89\,s for NPCS.

The encoding changes the times by factors that depend on the engine. Standard reification stores four statements per triple instead of two and matches a triple with its identifier by three triple patterns instead of one. Over the queries that all three encodings complete, standard reification is slower than RDF-star for every engine, method and size, by median factors from 1.02 to 3.94. Relative to RDF-star, named graphs take 0.74 to 0.93 times as long on Oxigraph for NPCS and the flat plan, 0.87 to 1.15 times on GraphDB and 1.06 to 1.74 times on Fuseki.

\begin{figure}[htbp]
\centering
\includegraphics[width=\columnwidth]{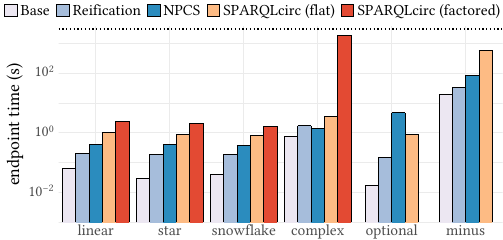}
\caption{Endpoint time per WatDiv class (109M triples).}
\Description{A logarithmic bar chart gives, for the linear, star, snowflake and complex templates of WatDiv and for the templates with OPTIONAL and MINUS, the endpoint time in seconds of Base, Reification, NPCS and the flat plan of SPARQLcirc on GraphDB with 109 million triples, together with the factored plan on the templates of basic graph patterns. A dotted line marks the budget.}
\label{fig:watdiv-class-100m}
\end{figure}

Figure~\ref{fig:watdiv-class-100m} repeats Figure~\ref{fig:watdiv} with 109 million triples.

\begin{figure*}[htbp]
\centering
\includegraphics[width=\textwidth]{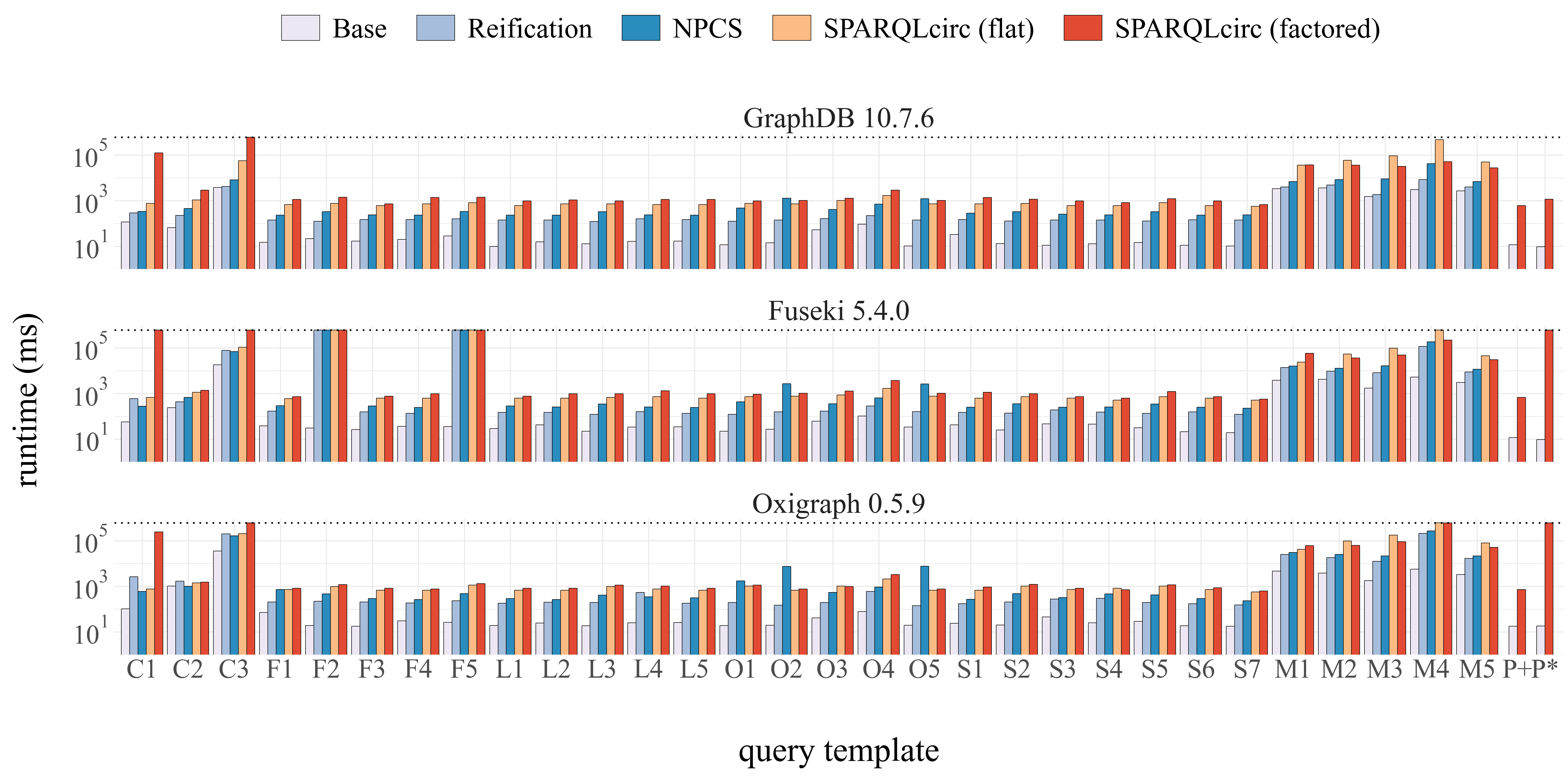}
\caption{Endpoint time per WatDiv template with 10.9 million triples and RDF-star annotations.}
\Description{Three logarithmic bar charts, for GraphDB, Fuseki and Oxigraph, give for every WatDiv template the endpoint time of the plain query, the reified query, NPCS and the flat and factored plans.}
\label{fig:watdiv-10m-rdfstar}
\end{figure*}

\begin{figure*}[htbp]
\centering
\includegraphics[width=\textwidth]{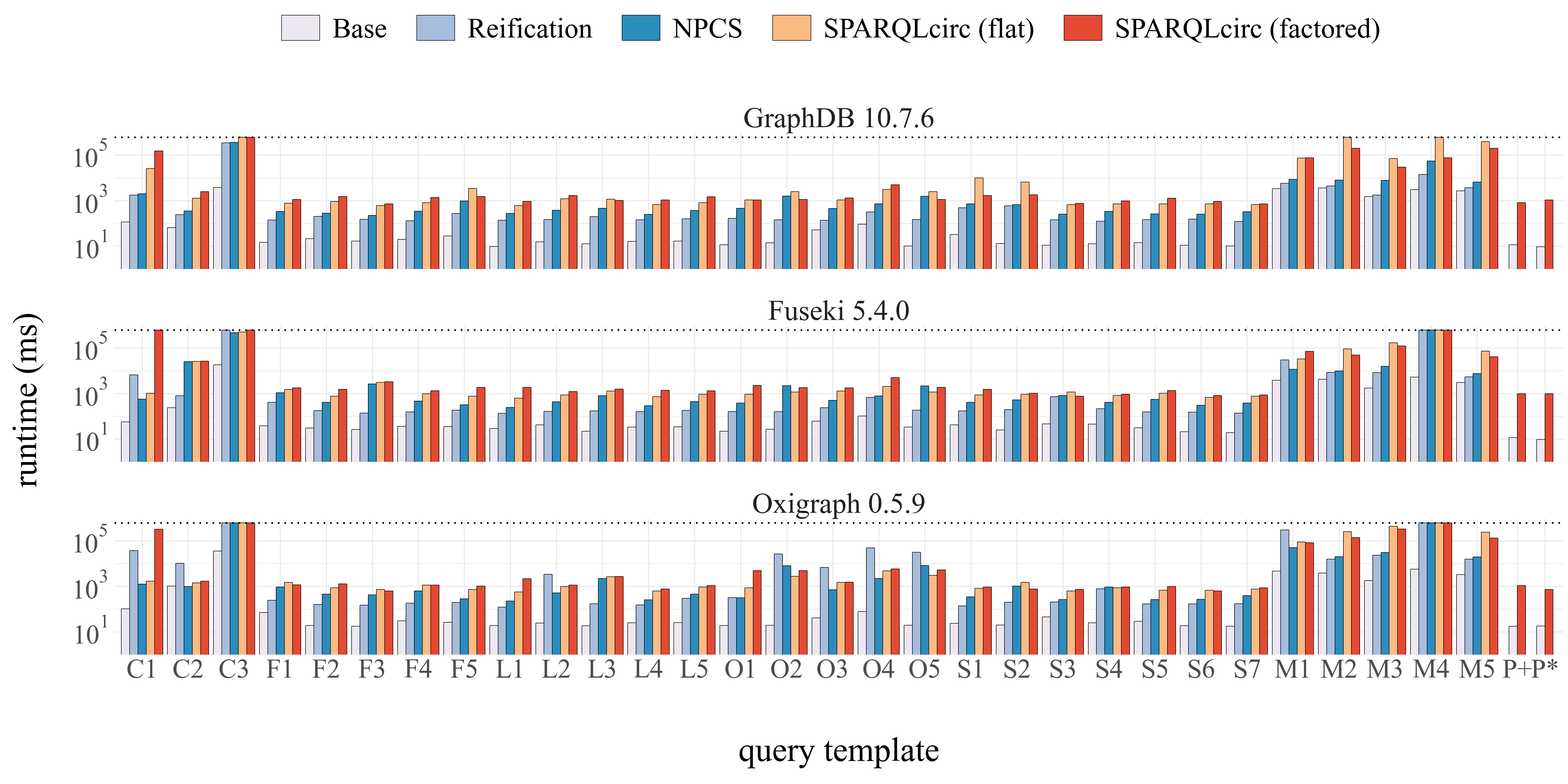}
\caption{Endpoint time per WatDiv template with 10.9 million triples and standard reification.}
\Description{Three logarithmic bar charts, for GraphDB, Fuseki and Oxigraph, give for every WatDiv template the endpoint time of the plain query, the reified query, NPCS and the flat and factored plans.}
\label{fig:watdiv-10m-standard}
\end{figure*}

\begin{figure*}[htbp]
\centering
\includegraphics[width=\textwidth]{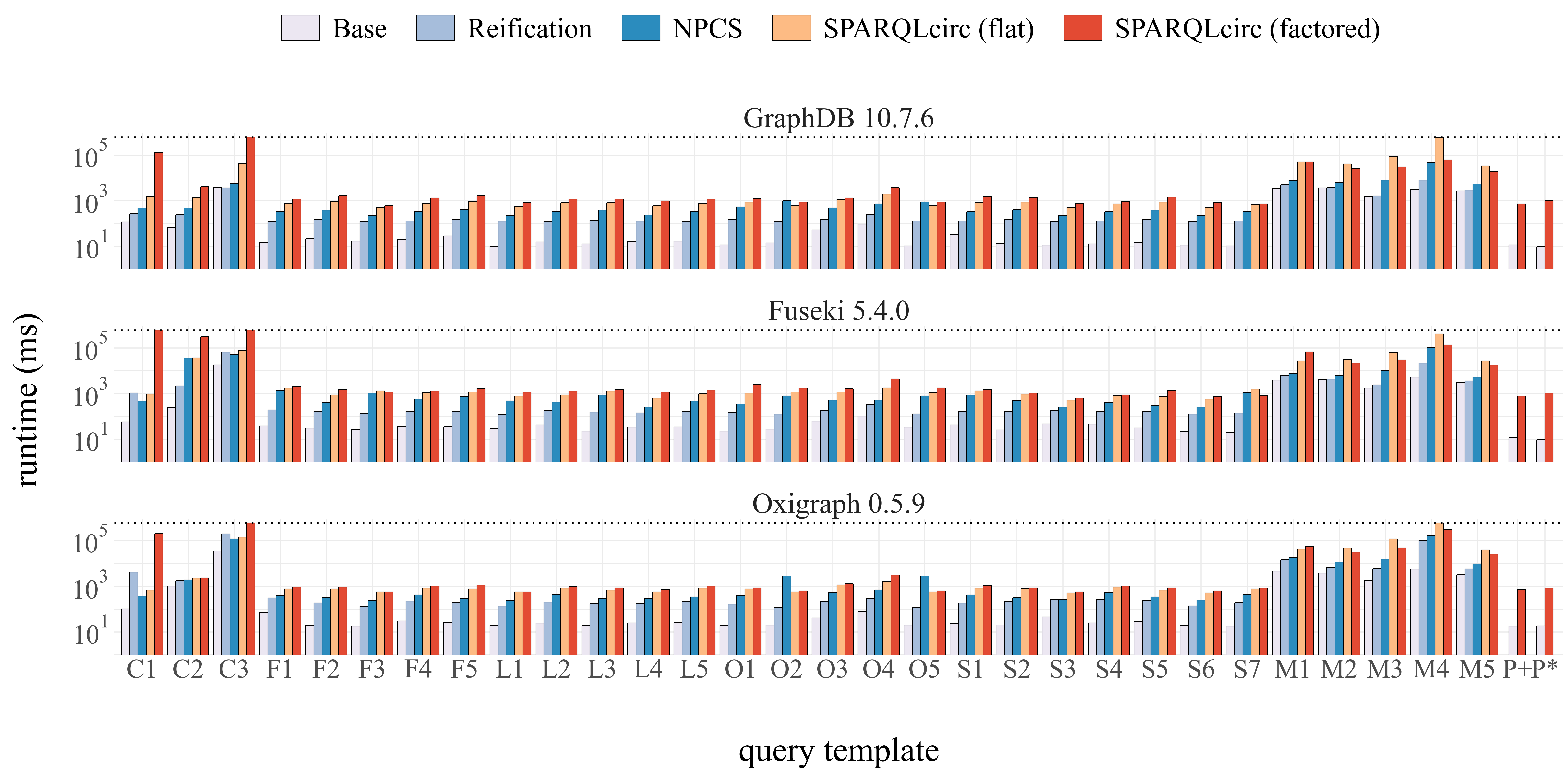}
\caption{Endpoint time per WatDiv template with 10.9 million triples and named graphs.}
\Description{Three logarithmic bar charts, for GraphDB, Fuseki and Oxigraph, give for every WatDiv template the endpoint time of the plain query, the reified query, NPCS and the flat and factored plans.}
\label{fig:watdiv-10m-namedgraph}
\end{figure*}

\begin{figure*}[htbp]
\centering
\includegraphics[width=\textwidth]{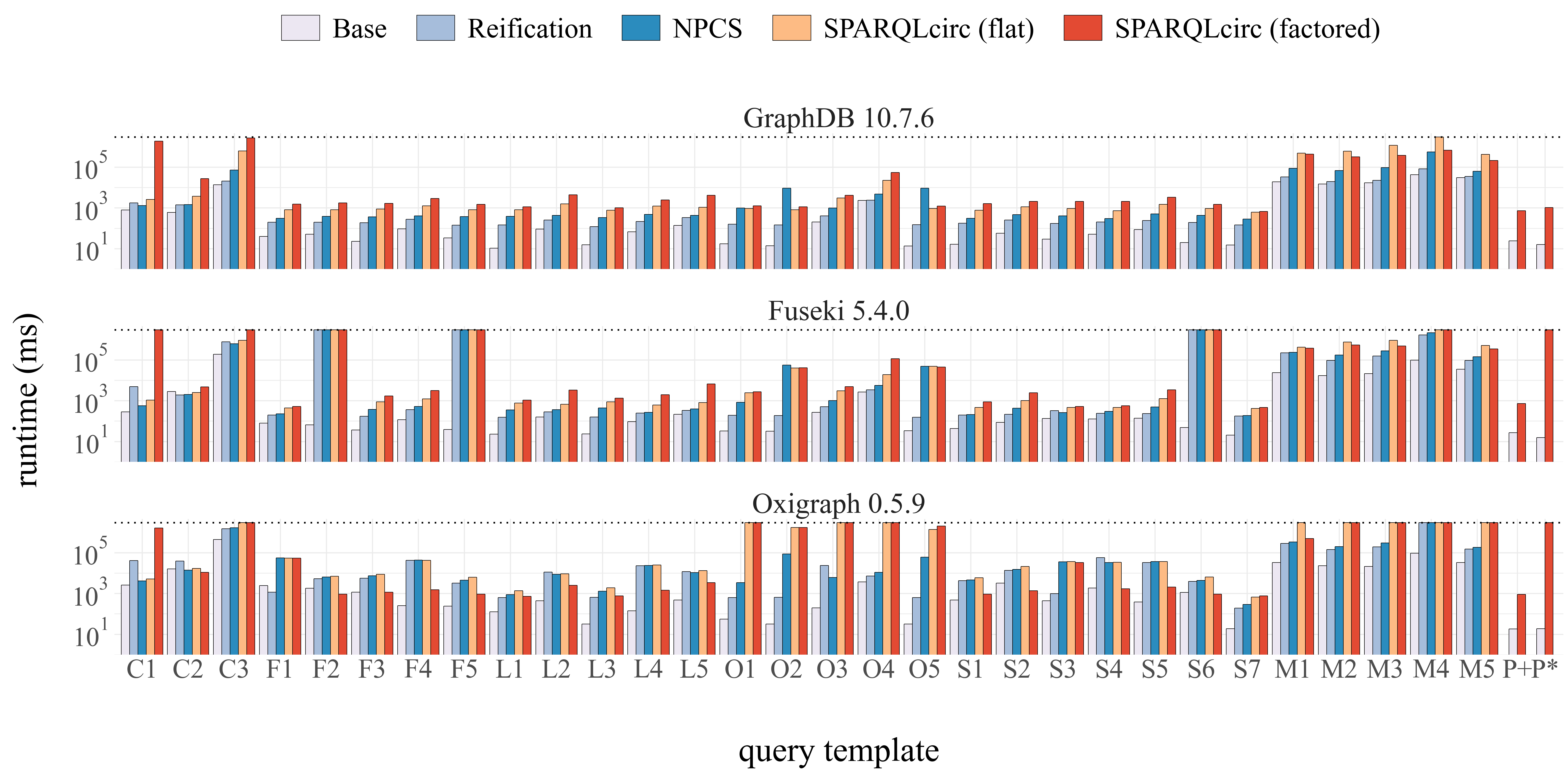}
\caption{Endpoint time per WatDiv template with 109 million triples and RDF-star annotations.}
\Description{Three logarithmic bar charts, for GraphDB, Fuseki and Oxigraph, give for every WatDiv template the endpoint time of the plain query, the reified query, NPCS and the flat and factored plans.}
\label{fig:watdiv-100m-rdfstar}
\end{figure*}

\begin{figure*}[htbp]
\centering
\includegraphics[width=\textwidth]{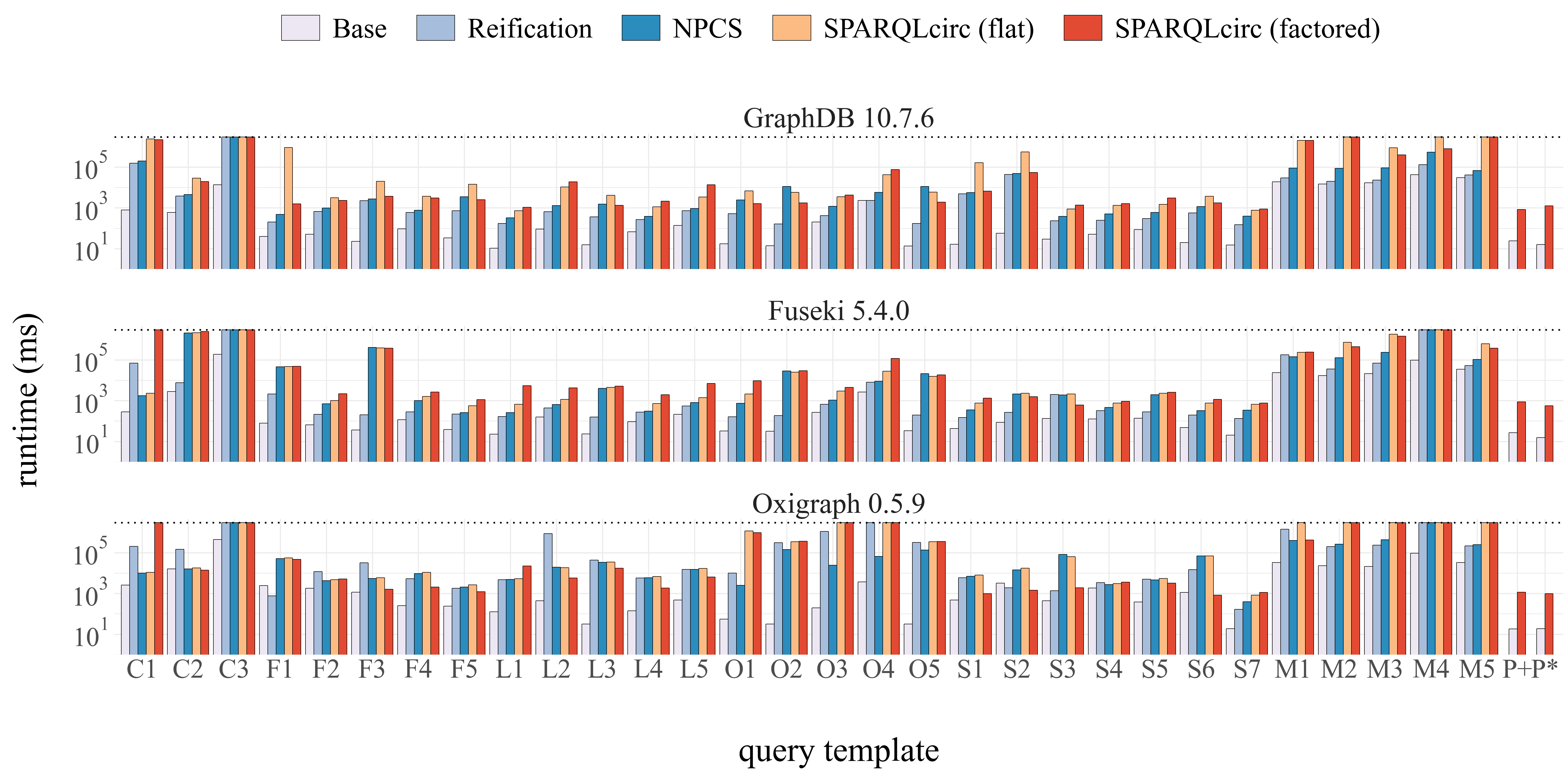}
\caption{Endpoint time per WatDiv template with 109 million triples and standard reification.}
\Description{Three logarithmic bar charts, for GraphDB, Fuseki and Oxigraph, give for every WatDiv template the endpoint time of the plain query, the reified query, NPCS and the flat and factored plans.}
\label{fig:watdiv-100m-standard}
\end{figure*}

\begin{figure*}[htbp]
\centering
\includegraphics[width=\textwidth]{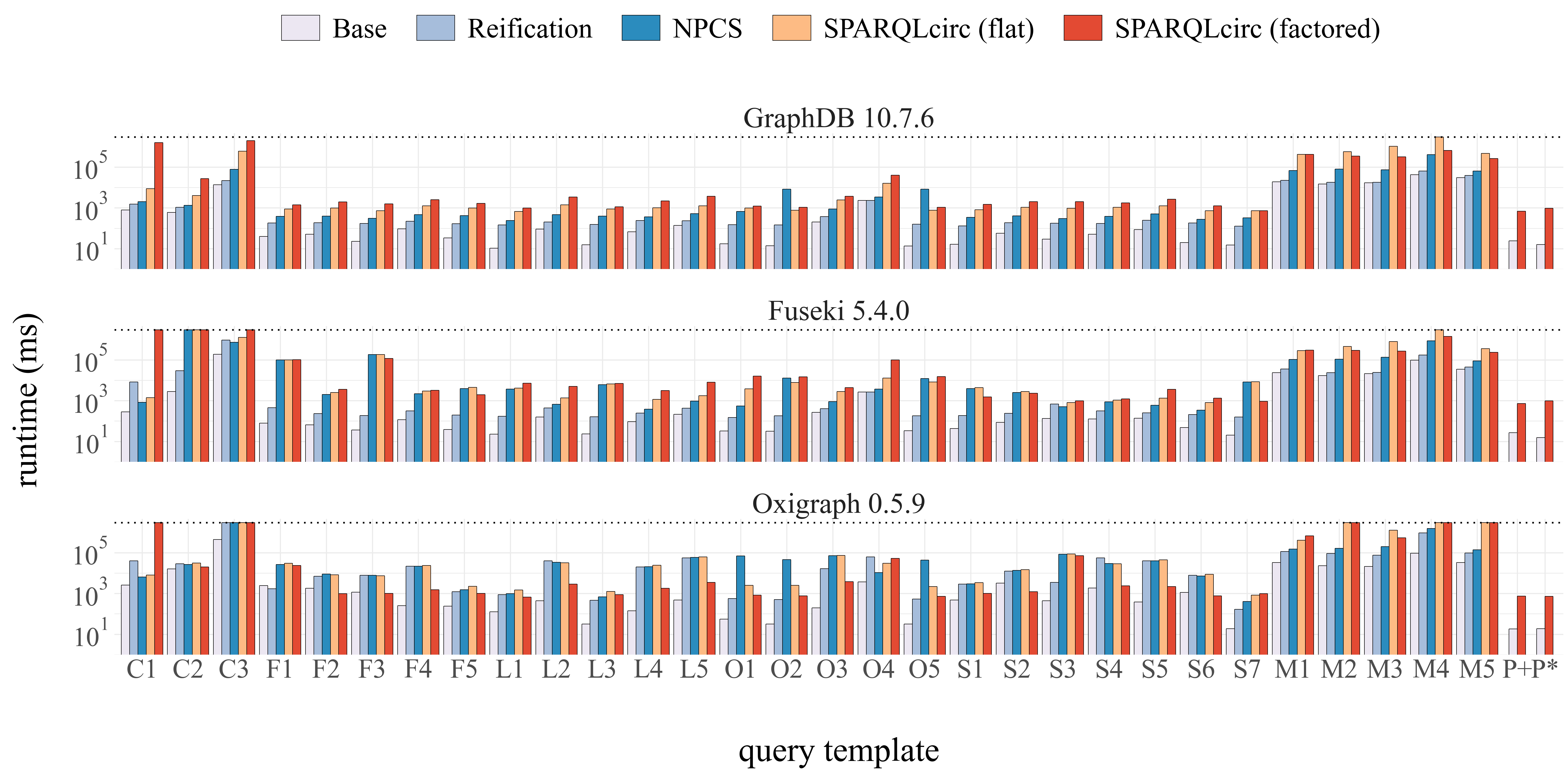}
\caption{Endpoint time per WatDiv template with 109 million triples and named graphs.}
\Description{Three logarithmic bar charts, for GraphDB, Fuseki and Oxigraph, give for every WatDiv template the endpoint time of the plain query, the reified query, NPCS and the flat and factored plans.}
\label{fig:watdiv-100m-namedgraph}
\end{figure*}

Figures~\ref{fig:watdiv-10m-rdfstar} to~\ref{fig:watdiv-100m-namedgraph} give the endpoint time of every template, where Base is the plain query and Reification the reified query. The two property-path templates P+ and P* run only the plain query and the plan of Section~\ref{sec:paths}.

\subsection{gMark}\label{app:exp:gmark}
At $k=6$, GraphDB runs with a heap of 128\,GiB instead of 48\,GiB, and \sys with 96\,GiB instead of 64\,GiB. Outside the endpoint, NPCS keeps about 0.1\,GiB, since it only stores the response, whereas the factored plan reaches 2 to 25\,GiB while it builds the circuit.

\subsection{Wikidata}\label{app:exp:wikidata}
The graph holds the statements of the full dump of 29~September 2026 without qualifiers and references in 8.87 billion triples.

\begin{figure*}[htbp]
\centering
\includegraphics[width=\textwidth]{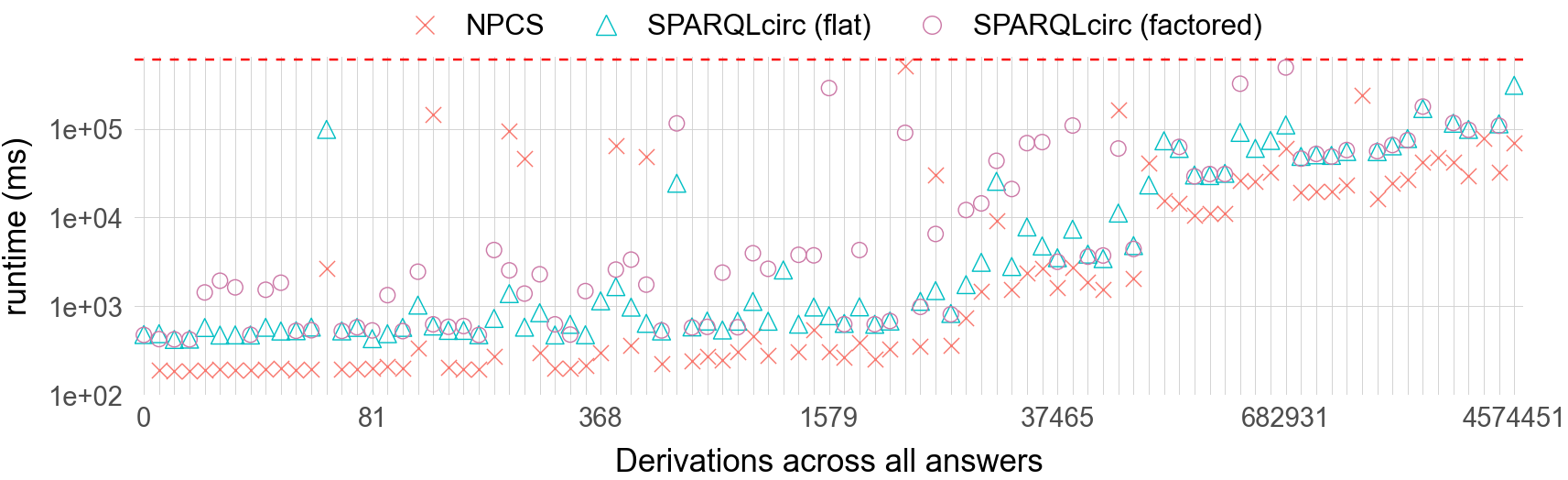}

\includegraphics[width=\textwidth]{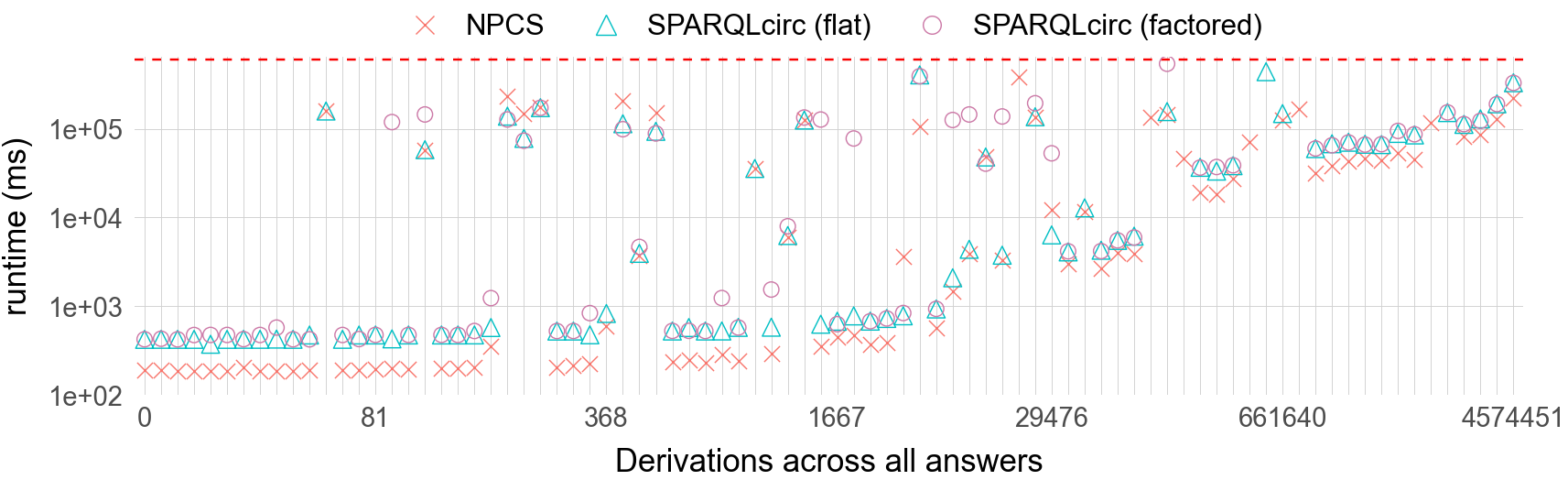}
\caption{Endpoint time of every Wikidata query against its number of derivations, on GraphDB (top) and Oxigraph (bottom).}
\Description{Two charts give, for every Wikidata query ordered by the number of derivations of all its answers, the endpoint time of NPCS and of the flat and factored plans on a logarithmic axis, on GraphDB and on Oxigraph. A dashed line marks the budget.}
\label{fig:wikidata-derivations}
\end{figure*}

\begin{figure*}[htbp]
\centering
\includegraphics[width=\textwidth]{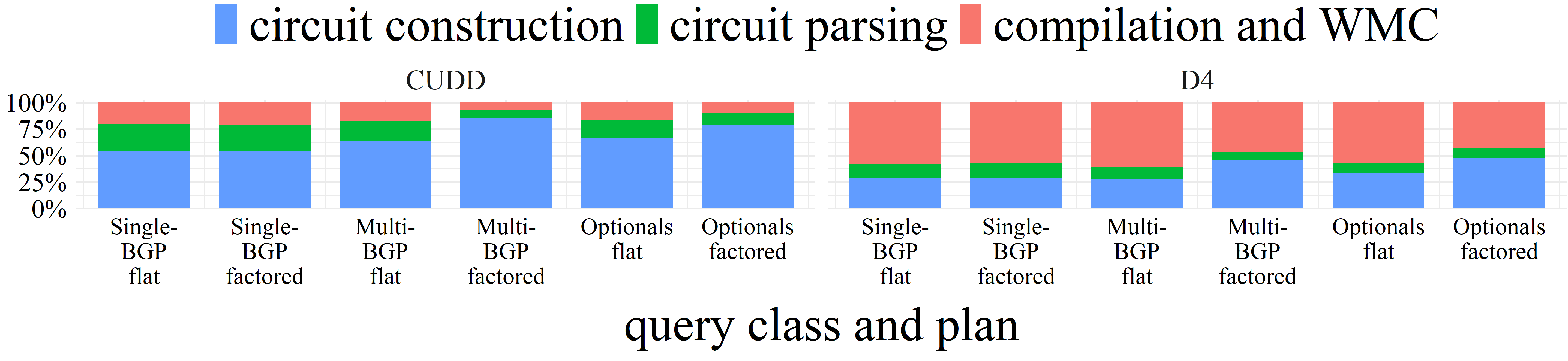}

\includegraphics[width=\textwidth]{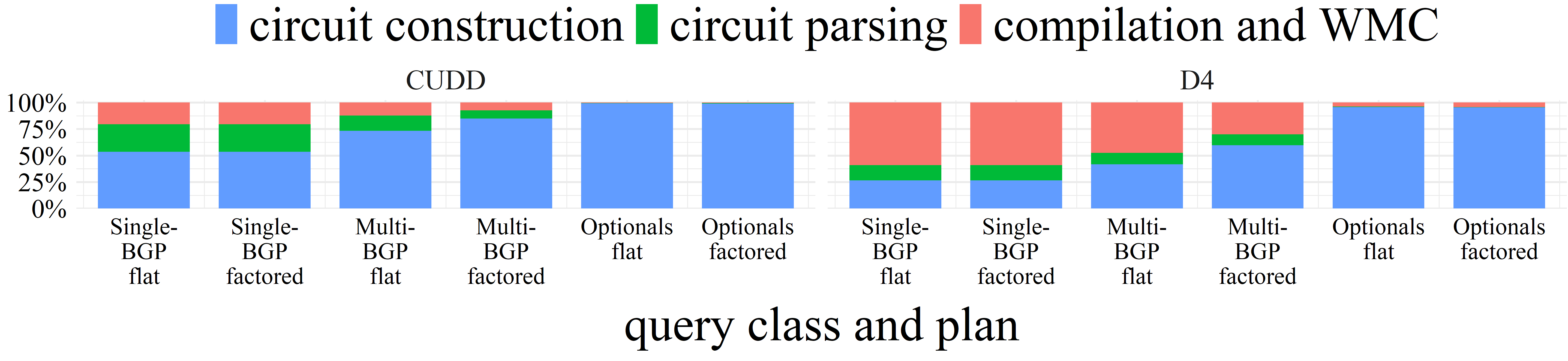}
\caption{Share of the stages in the end-to-end time of \sys on Wikidata, on GraphDB (top) and Oxigraph (bottom).}
\Description{Stacked bar charts give, for every query class and plan and for CUDD and d4, the share of circuit construction, circuit parsing and compilation with weighted model counting in the end-to-end time.}
\label{fig:wikidata-stages}
\end{figure*}

At the endpoint, NPCS completes on GraphDB 42 queries with one triple pattern, 32 with several and 13 with \OPT{}, the flat plan 42, 30 and 15, and the factored plan 41, 25 and 14. On Oxigraph, NPCS completes 82 queries, the flat plan 77 and the factored plan 71, and the flat plan completes 5 of the queries with \OPT{} against 10 for NPCS. At the median, the flat plan takes 2.5 times as long as NPCS on GraphDB and 1.7 times on Oxigraph (Figure~\ref{fig:wikidata-derivations}), and the factored plan 2.8 and 2.2 times. Figure~\ref{fig:wikidata-derivations} gives the endpoint time of every query, which ends before NPCS parses and merges its provenance expressions. NPCS is therefore faster at the endpoint on almost every query, and the flat plan overtakes it only end to end on queries with many derivations (Table~\ref{tab:wikidata}). End to end with CUDD, the flat plan returns all probabilities for 87 queries on GraphDB and 77 on Oxigraph, and NPCS for 82 and 80. If NPCS compiles the expression of every answer separately instead of merging them, it completes 66 and 64 queries, and the flat plan is faster on 49 of the 62 queries that both complete on GraphDB, by 2.7 times at the median. With d4, compilation and weighted model counting take 54 to 68\% of the end-to-end time of the flat plan, against about a fifth with CUDD (Figure~\ref{fig:wikidata-stages}), and the flat plan and NPCS take about the same time end to end.

On GraphDB, the flat plan is more than ten times faster than NPCS on seven queries with \OPT{}, with 2.1\,s against 67\,s at the median, although five of them have fewer than 500 derivations. On Oxigraph, the flat plan is at most twice as fast with \OPT{} and exceeds the budget on four of these queries that NPCS completes.

\subsection{TPC-H}\label{app:exp:provsql}
TPC-H~\cite{tpch} at nine scale factors from 0.01 to 1 becomes 1.2 to 123 million triples under the mapping of SPARQLprov~\cite{hernandez2021sparqlprov}, in which the IRI of a row identifies its triples (Section~\ref{sup:builder}). ProvSQL~1.11.0-dev~\cite{senellart2018provsql} in PostgreSQL~18.4 runs on a relational copy.

The time of \sys grows about linearly with the size of TPC-H, as does that of ProvSQL. The flat plan completes every scale factor on six of the twelve templates on GraphDB and on seven on Oxigraph, and ProvSQL on eleven. On these templates, the time of the flat plan grows with the scale factor at a median exponent of 0.78 on GraphDB and 0.95 on Oxigraph, against 0.93 and 0.90 for ProvSQL. ProvSQL is faster in absolute terms, by 20 times at the median for the flat plan on GraphDB, because the direct mapping turns every attribute of a row into a triple, so a template joins more triple patterns than ProvSQL joins tables.

\begin{figure*}[htbp]
  \centering
  \includegraphics[width=\textwidth]{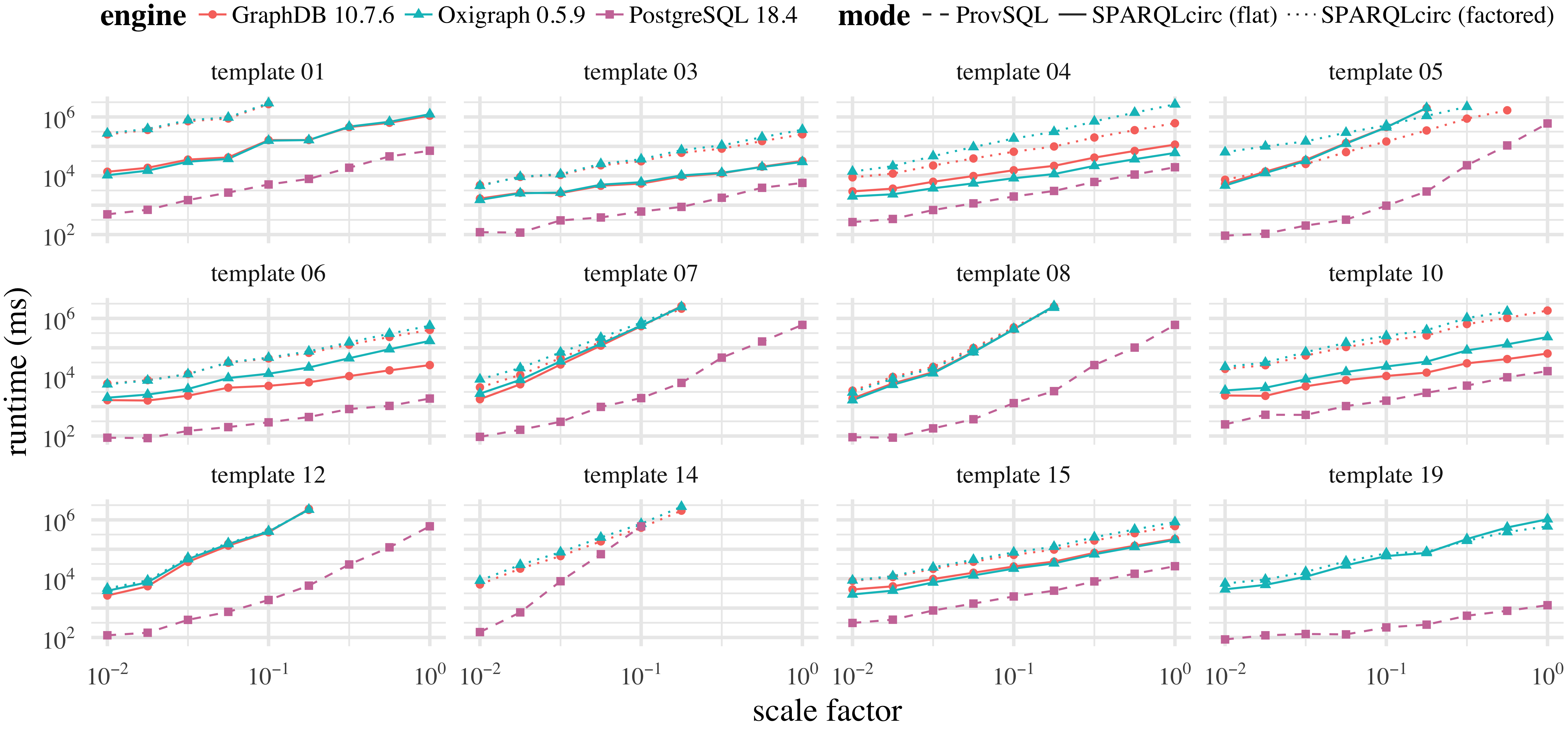}
  \caption{Exact PQE end to end on TPC-H, one panel per template. A line stops short where the
  budget was exhausted.}
  \Description{Twelve logarithmic line charts, one per TPC-H template, plot end-to-end latency against the scale factor for ProvSQL and for the flat and factored plans on GraphDB and Oxigraph.}
  \label{fig:tpch}
\end{figure*}
Figure~\ref{fig:tpch} gives the end-to-end time on every template. On GraphDB, ProvSQL is faster than the flat plan on every template, by 5.7 to 288 times at the median of a template, and the factored plan takes 121 times as long as ProvSQL at the median.

\clearpage
\clearpage
\section{Notation}\label{app:notation}
\begin{table}[H]
\caption{Notation.}
\label{tab:notation}
\footnotesize
\begin{tabular}{@{}l@{\hspace{6pt}}p{0.6\columnwidth}@{}}
\toprule
Symbol & Meaning\\
\midrule
$\mathbf{I}$, $\mathbf{B}$, $\mathbf{L}$, $\mathbf{V}$, $\mathbf{T}$ & IRIs, blank nodes, literals, variables, RDF terms\\
\frag & SPARQL fragment (Section~\ref{sec:prelim})\\
$G$ & {RDF graph underlying the probabilistic model}\\
{$t$, $\tp$, $P$, $Q$} & {triple, triple pattern, pattern, query}\\
$W$, $x_t$ & possible world $W\subseteq G$, variable of $t\in G$\\
$\mu$, $\nu$, $\sim$ & solution mappings and compatibility\\
{$\sem{P}{G}$, $\supp$} & {multiset evaluation and its support}\\
$Y$ & projected variables of a query\\
{$\mu^C_{s,v}$} & {mapping of closure atom $C$ for $s$ and $v$}\\
$\val{W}$ & valuation that makes $x_t$ true exactly when $t\in W$\\
{$X$} & {variables $\{x_t\mid t\in G\}$}\\
$\BProv{P}{\mu}$ & Boolean provenance of $\mu$ for pattern $P$\\
$\Pr(x_t)$, $\Pr(\mu)$, $\mathrm{WMC}$ & probability of $t$, marginal probability of $\mu$, weighted model count\\
{$G_{\mathrm{reif}}$} & {reified graph that the endpoint stores}\\
$\oplus$, $\otimes$, $\ominus$ & sum, product and difference gates\\
{$\mathit{ch}$, $\mathit{pos}$, $\mathit{neg}$} & {children, minuends and subtrahend of a gate}\\
{$\mathrm{root}(\mu)$} & {root of the mapping $\mu$}\\
$\dec{g}$ & Boolean function of gate $g$ ({Section~\ref{sec:prelim}})\\
{$g_\bot$, $g_\top$} & {childless gates for $\bot$ and $\top$}\\
{$\tau$, $\theta(P)$} & {term encoding, serialization of $P$}\\
{$\bar v_P$, $\bar Y$} & {scope and projection tuples in lexicographic order}\\
$\delta$, $\delta_{128}$ & lower-cased SHA-256 digest, its first 128 bits\\
\texttt{KEY\_}$\ast$, $h$, $\operatorname{sort}$, $\hat\tau$, $L$ & key expressions, their parts and the letter of a sum or difference gate (Definition~\ref{def:keys})\\
$\rewr(P)$, $g_P$ & plan of \CONSTRUCT{} queries and gate variable of $P$\\
$q_P$, $q_{P,i}$, $q^-_P$, $q^+_P$ & queries of $P$ (Definition~\ref{def:gamma})\\
$\reifb{\tp}{?t}$ & reification of $\tp$ with identifier variable $?t$\\
$\rwp{P}{g_P}$ & rewritten pattern of $P$ that binds $g_P$\\
$\cand{P}{\mu}$, $\Phi_P(\mu)$ & gates bound by the rows of $\rwp{P}{g_P}$ for $\mu$, disjunction of their functions\\
$s_P$ & subtrahend gate of a difference $P$\\
$m$, $Z$, $F_j$ & number of triple patterns, variables of $Q$ outside $Y$, base factor of $\tp_j$\\
$\mathcal A_z$, $H_z$ & factors whose scopes contain $z$, the factor that replaces them\\
$A_z(\nu)$, $\adom(G)$ & values of $z$ at which every factor of $\mathcal A_z$ is defined, terms of $G$\\
$U_z$, $w$ & scope of $H_z$, induced width\\
$\mathcal I_C$, $S_C$ & input relation and sources of a closure atom $C$\\
$g_C(s,v)$ & gate of $\mu^C_{s,v}$\\
$e$, $e^\diamond$, $s$, $v$ & closure-free expression, closure ($\diamond\in\{+,*\}$), source, target\\
$D_G$, $C_e$ & subjects and objects of $G$, closure-free circuit of $e$ (Section~\ref{app:full-paths})\\
$\relpath{e}{G}$, $\relpath{e^+}{G}$ & relation of $e$ on $G$ and its transitive closure\\
$\Delta$ & zero-length pairs of a closure $e^*$ (Section~\ref{sec:prelim})\\
$C=(\alpha,e^\diamond,?y)$ & closure atom with source $\alpha$ and target variable $?y$\\
$b^e_{uv}$ & gate that is true in $W$ exactly when $(u,v)\in\relpath{e}{W}$\\
$V_s$, $E_s$, $n$ & nodes and $e$-steps reachable from $s$, $n=|V_s|$\\
$r_i(v)$, $R(v)$, $\lambda$ & {gate for length-$i$ walks, closure gate of $v$, last level}\\
$o_s$ & zero-length gate: sum of the leaves of the triples with subject or object $s$\\
$J$, $J_t$, $y_i$ & identifiers, identifiers of $t$, variable of identifier $i$ (Appendix~\ref{app:builder})\\
$\sigma$, $W_\omega$ & substitution of $x_t$ by $\bigvee_{i\in J_t}y_i$, world of the identifiers $\omega$\\
\bottomrule
\end{tabular}
\end{table}

\end{document}